\documentclass[twocolumn]{aa}
\usepackage{graphicx}
\usepackage{txfonts}
\begin{document}
   \title{The ratio of $N$(C$^{18}$O) and $A_\mathrm{V}$ in Chamaeleon I and III-B}

   \subtitle{Using 2MASS and SEST}

   \author{Kainulainen, J. \and Lehtinen K. \and Harju, J.
             }
 
   \offprints{Kainulainen, J. (jkainula@astro.helsinki.fi)}

   \institute{Observatory, P.O. Box 14, SF-00014 University of Helsinki}

   \date{Received <date> / Accepted <date>}

   \abstract{We investigate the relationship between the C$^{18}$O column density and the visual extinction in Chamaeleon I and in a part of the Chamaeleon III molecular cloud. The C$^{18}$O column densities, $N$(C$^{18}$O), are calculated from $J=1-0$ rotational line data observed with the SEST telescope. The visual extinctions, $A_\mathrm{V}$, are derived using $JHK$ photometry from the 2MASS survey and the NICER color excess technique. In contrast with the previous results of Hayakawa et al. (\cite{hayakawa01}), we find that the average $N$(C$^{18}$O)$/A_\mathrm{V}$ ratios are similar in Cha I and Cha III, and lie close to values derived for other clouds, i.e. $N$(C$^{18}$O) $\approx 2 \times 10^{14} \mathrm{ cm}^{-2} ( A_\mathrm{V} - 2 )$. We find, however, clear deviations from this average relationship towards individual clumps. Larger than average $N$(C$^{18}$O)$/A_\mathrm{V}$ ratios can be found in clumps associated with the active star forming region in the northern part of Cha I. On the other hand, some regions in the relatively quiescent southern part of Cha I show smaller than average $N$(C$^{18}$O)$/A_\mathrm{V}$ ratios and also very shallow proportionality between $N$(C$^{18}$O) and $A_\mathrm{V}$. The shallow proportionality suggests that C$^{18}$O is heavily depleted in these regions. As the degree of depletion is proportional to the gas density, these regions probably contain very dense, cold cores, which do not stand out in CO mappings. A comparison with the dust temperature map derived from the ISO data shows that the most prominent of the potentially depleted cores indeed coincides with a dust temperature minimum. It seems therefore feasible to use $N$(C$^{18}$O) and $A_\mathrm{V}$ data together for identifying cold, dense cores in large scale mappings.

               }
 
   \maketitle
%

\section{Introduction}

The most common constituent of interstellar molecular clouds is molecular hydrogen, H$_2$, which composes about 70 \% of their total masses. However, direct determination of the H$_2$ column density is not generally possible as the molecule does not emit observable line radiation at low temperatures. The distribution of molecular gas, and furthermore the mass, in molecular clouds is therefore usually derived by observing the isotopes of carbon monoxide, namely: $^{12}$CO, $^{13}$CO, C$^{17}$O and C$^{18}$O. The most common way to derive the H$_2$ distribution is to observe a single transition of C$^{18}$O, usually $J=1-0$, calculate the C$^{18}$O column density, $N$(C$^{18}$O), by assuming the local thermodynamical equilibrium (LTE) and use a constant, ``canonical'', $N$(H$_2)$ / $N$(C$^{18}$O) ratio in the conversion to H$_2$ column density. This method is facilitated by the facts that the transitions of C$^{18}$O are usually optically thin, and easily thermalized because of the relatively low dipole moment of molecule.

There is, however, no theoretical reason for the constancy of the $N$(H$_2$) / $N$(C$^{18}$O) ratio in the varying conditions of the interstellar matter. In the cold and dense environment, the gas molecules will be depleted on the surfaces of the dust grains, and thus become absent from the total gaseous reservoir (e.g. Little et al. \cite{little79}; Burke \& Hollenbach \cite{burke83}). This phenomenon is most obvious in dense and dark globules, in which abundances of molecules such as CO and CS can be decreased by one or two orders of magnitude towards the globule center (Kuiper et al. \cite{kuiper96}; Willacy et al. \cite{willacy98}; Kramer et al. \cite{kramer99}; Alves et al. \cite{alves99}; Caselli et al. \cite{caselli99}; Bergin et al. \cite{bergin01}; Jessop \& Ward-Thompson \cite{jessop01}; Hotzel et al. \cite{hotzel01}; Hotzel et al. \cite{hotzel02}; Tafalla et al. \cite{tafalla02}). On the other hand, molecules frozen on the grain surfaces can be evaporated because of heating by energetic particles or radiation. Especially, heating due to radiation and schocks in the vicinity of vigorous star formation should alter the level of depletion (e.g. Williams \cite{williams85}; Charnley et al. \cite{charnley88}). Williams (\cite{williams85}) concludes that variations can occur from cloud to cloud, and even within the different regions of the same cloud reflecting the local star formation activity.

The dust component of interstellar matter provides another type of mass tracer in molecular clouds, as the dust particles absorb and scatter the light of the stars behind the cloud. The extinction of light, which is directly proportional to intervening dust column density, is very high at optical wavelengths and the dense cloud cores are totally opaque. However, in the near-infrared the extinction is only about one tenth of that in the visible and cloud cores are more transparent. It is therefore viable to derive $N$(H$_2)$ and the overall mass distribution in a molecular cloud with the aid of near-infrared observations.

The $N$(C$^{18}$O) / A$_\mathrm{V}$ ratio has been determined for several clouds, beginning with the work of Encrenaz et al. (\cite{encrenaz75}) in Rho Ophiuchi. Since then, the ratio has been examined by Frerking et al. (\cite{frerking82}) in Taurus and $\rho$ Ophiuchi, Harjunp\"a\"a et al. (\cite{harjunpaa96}) in Chamaeleon I, R Corona Australis and Coalsack, Hayakawa et al. (\cite{hayakawa99, hayakawa01}) in the Chamaeleon complex, and by Lada et al. (\cite{lada94}), Alves et al. (\cite{alves99}), Juvela et al. (\cite{juvela97, juvela02}) and Harjunp\"a\"a et al. (\cite{harjunpaa04}) in several small globules. These data, reviewed and compared in more detail by Harjunp\"a\"a et al. (\cite{harjunpaa04}), show some evidence supporting the correlation of $N$(C$^{18}$O) / A$_\mathrm{V}$ with star forming activity. 

We present the derivation of the $N$(C$^{18}$O) / $A_\mathrm{V}$ ratio in Chamaeleon I, and in selected region of Chamaeleon III, with emphasis on the possible difference between star forming and non-star forming regions. The present study is both deeper in extinction and more sensitive in C$^{18}$O than the previous large scale studies conducted in Chamaeleon region . In \S\ref{sec_chamaeleon-cloud-complex} the Chamaeleon cloud complex is briefly introduced. In \S\ref{sec_observations} we present the CO observations and NIR data used in the study. The relations between the derived quantities, $N$(C$^{18}$O) and $A_\mathrm{V}$, are given in \S\ref{sec_correlations}, and these results are discussed in \S\ref{sec_discussion}. In \S\ref{sec_conclusions} we summarize the results.
 

\section{Chamaeleon cloud complex}
\label{sec_chamaeleon-cloud-complex}

\begin{figure*}[p!]
   \centering
   \includegraphics[width=1.25\columnwidth]{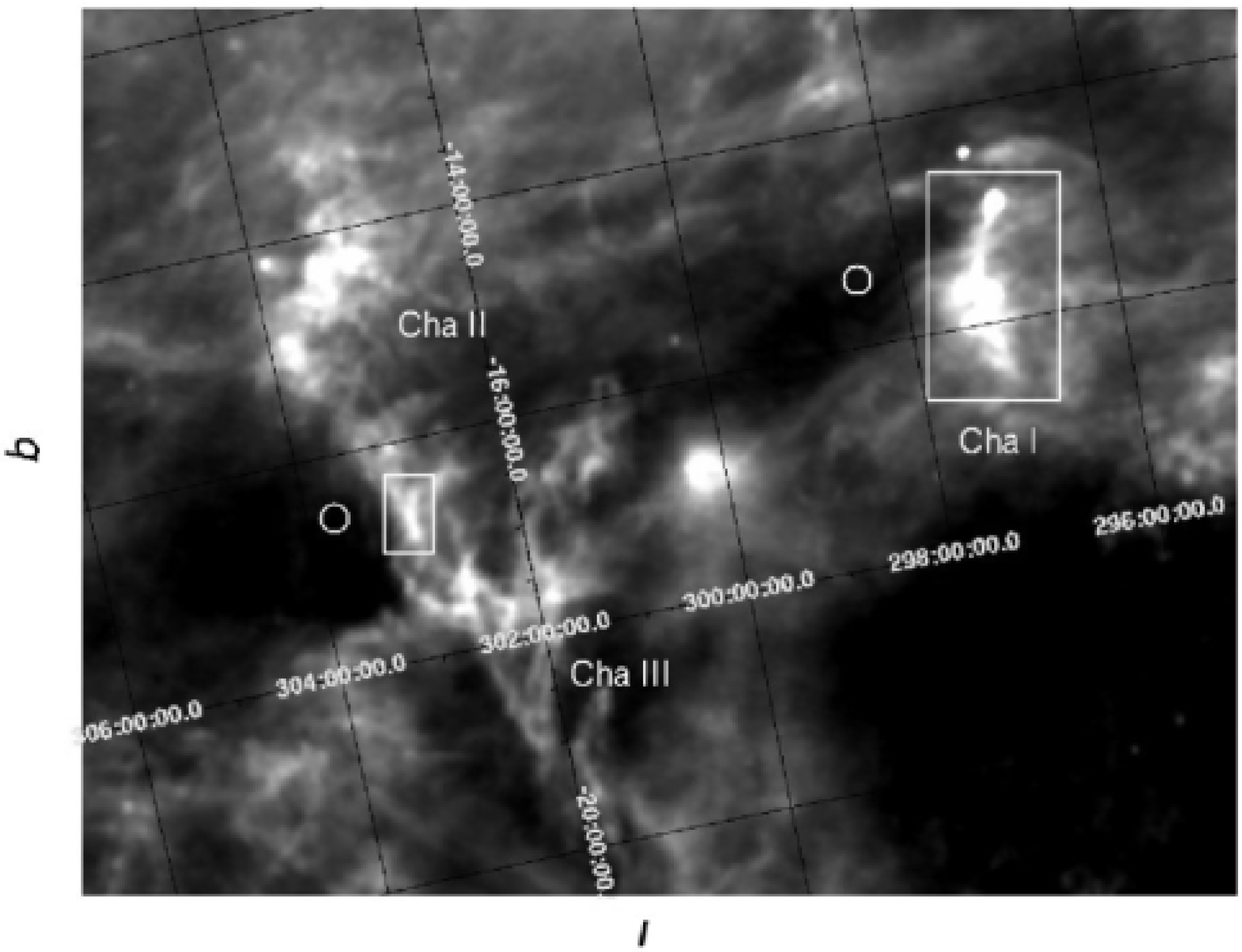}
      \caption{IRAS 100 $\mu$m emission map of the Chamaeleon cloud complex.
Chamaeleon I and III-B are marked with rectangles, and positions
of the reference fields used in the extinction calculation are marked with circles.}
         \label{IRAS100}
   \end{figure*}

The Chamaeleon cloud complex is an extensive concentration of molecular clouds covering dozens of square degrees on the sky. The complex has been studied at several wavelengths and it is proven very useful for studying low-mass star formation due to its close proximity (150 pc from the Hipparcos satellite data. Knude \& H\o g \cite{knude98}), somewhat isolated location below the galactic plane ($b \approx -16^o$), and the undisturbed view along the line of sight. Among the three large clouds in the complex, Chamaeleon I (Cha I) is the most active stellar factory, with about 200 identified Young Stellar Objects (YSOs) within its boundaries. In contrast, Chamaeleon III (Cha III) is a quiescent cloud, without current evidence for star formation. This variety makes the complex particularly interesting. Fig. \ref{IRAS100} shows the IRAS 100 $\mu$m emission map of the complex, with the regions investigated in this study denoted.

On a large scale, Cha I has not been mapped at many molecular line frequencies. Observations have been made in the emission lines of H$_2$CO and OH (Toriseva et al. 1985), $^{12}$CO, $^{13}$CO and C$^{18}$O (Toriseva et al. \cite{toriseva90}; Mizuno et al. \cite{mizuno98, mizuno99, mizuno01}; Haikala et al. \cite{haikala05}). The infrared survey using the DENIS (DEep Near-Infrared Survey of the southern sky) photometry was done towards Cha I by Cambresy et al. (\cite{cambresy98}) to reveal young stellar objects. G\'omez \& Kenyon (\cite{gomez01}) also used DENIS to search for YSOs and to derive the extinction law in Cha I. Extinction maps of Cha I have been previously derived by Toriseva et al. (\cite{toriseva85b}) and Cambresy et al. (\cite{cambresy97}). Several studies have used different techniques and instruments including the Hubble Space Telescope, the Very Large Telescope and the Infrared Space Observatory (ISO) to identify and characterize individual objects on a smaller spatial scale (e.g. Oasa, Tamura and Sugitani \cite{oasa99}; Lehtinen et al. \cite{lehtinen01}; Persi et al. \cite{persi01}; Kenyon \& G\'omez \cite{kenyon01}; Carpenter \cite{carpenter02}; Stelzer et al. \cite{stelzer04}).

The previous large scale studies of Chamaeleon I (Mizuno et al. \cite{mizuno99} and T\'oth et al. \cite{toth00}; Haikala et al. \cite{haikala05}) all use different naming convention for the clumps or cores identified in the cloud. Throughout this paper we use the naming convention of Haikala et al. (\cite{haikala05}) whenever applicable. See the Fig. \ref{fig_co-map-chaI} for the positions of some of these cores.

Cha III is similar in size to Cha I but has a lower column density: the IRAS 100 $\mu$m maximum surface brightness is only about 20 \% of that of Cha I. Cha III is an example of a filamentary structure with a possible contribution of magnetic fields creating a wavy, serpentine-like architecture (Gahm et al. \cite{gahm02}). The cloud has been observed in $^{13}$CO and C$^{18}$O by Mizuno et al. (\cite{mizuno99}), and in part by Gahm et al. (\cite{gahm02}). Following the labelling of Gahm et al. (\cite{gahm02}), an area denoted as ChaIII-B has been selected for this study. It represents one filament of Cha III and is prominent in the IRAS $\Delta I_{100}=I(100 \mu $m$) - 5 I(60 \mu $m$)$ map, containing probably the most dense matter in Cha III. The quantity  $\Delta I_{100}$ has been found to have a very good correlation with $^{13}$CO, representing the cold dust emission in molecular clouds (Laureijs et al. \cite{laureijs95}). Cha III-B, among other filaments, has been studied in more detail by e.g. Gahm et al. (\cite{gahm02})

The $N$(C$^{18}$O)$/ A_\mathrm{V}$ ratio has been examined in the Chamaeleon clouds by Hayakawa et al. (\cite{hayakawa99, hayakawa01}) with J-band photometric data from DENIS and C$^{18}$O observations from the NANTEN radiotelescope (2.7$'$ FWHM beam). Hayakawa et al. (\cite{hayakawa01}) studied the ratio also in different parts within Cha I. Harjunp\"a\"a and Mattila (\cite{harjunpaa96}) derived the $N$(C$^{18}$O) / $A_\mathrm{V}$ ratio for about 20 individual lines of sight across Cha I using the SEST telescope and near-IR extintion values from literature. The results of the present study are compared with these results in section \ref{sec_discussion}.


\section{Observations}

\label{sec_observations}

\subsection{CO data}

\label{sec_co-observations}

The C$^{18}$O (J=1-0) observations of Cha I and Cha III at 109 GHz were made during several observing runs between 1988 and 1996 with the SEST telescope on La Silla in Chile. The data have been published by Gahm et al. (\cite{gahm02}, Cha III), and by Haikala et al. (\cite{haikala05}, Cha I, H05 hereafter). The observations and the instrumentation are described in detail in these articles. The Cha I map covers completely the high column density region in the cloud, about 30$'$ $\times$ 90$'$ or 0.6 square degrees in size, with a 1$'$ spacing between the observed positions. The SEST beam size at this frequency is 45$''$. The Cha III map used in the present paper covers a filament called Cha III-B in Gahm et al. (\cite{gahm02}), with dimensions 10$'$ $\times$ 30$'$. The map spacing in the observations alternates between 30$''$ and 1$'$. We have used the original data provided by the authors of the two C$^{18}$O papers. The data have been smoothed to correspond to the resolution 2$'$ achieved in the extinction maps from the 2MASS data (section \ref{sec_visual-extinctions}).

\begin{figure*}[p!]
   \centering
   \includegraphics[width=0.99\columnwidth]{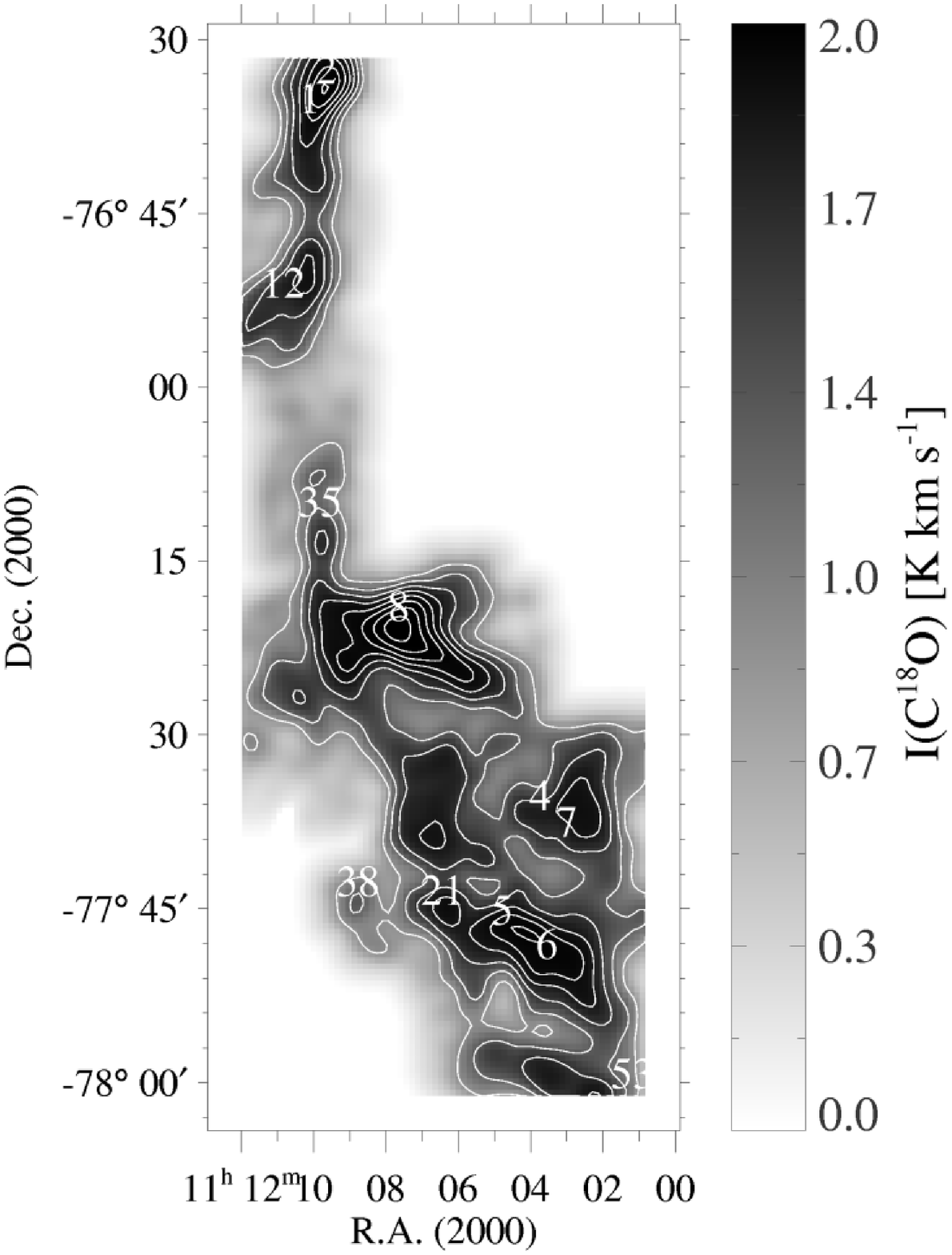} \includegraphics[width=0.99\columnwidth]{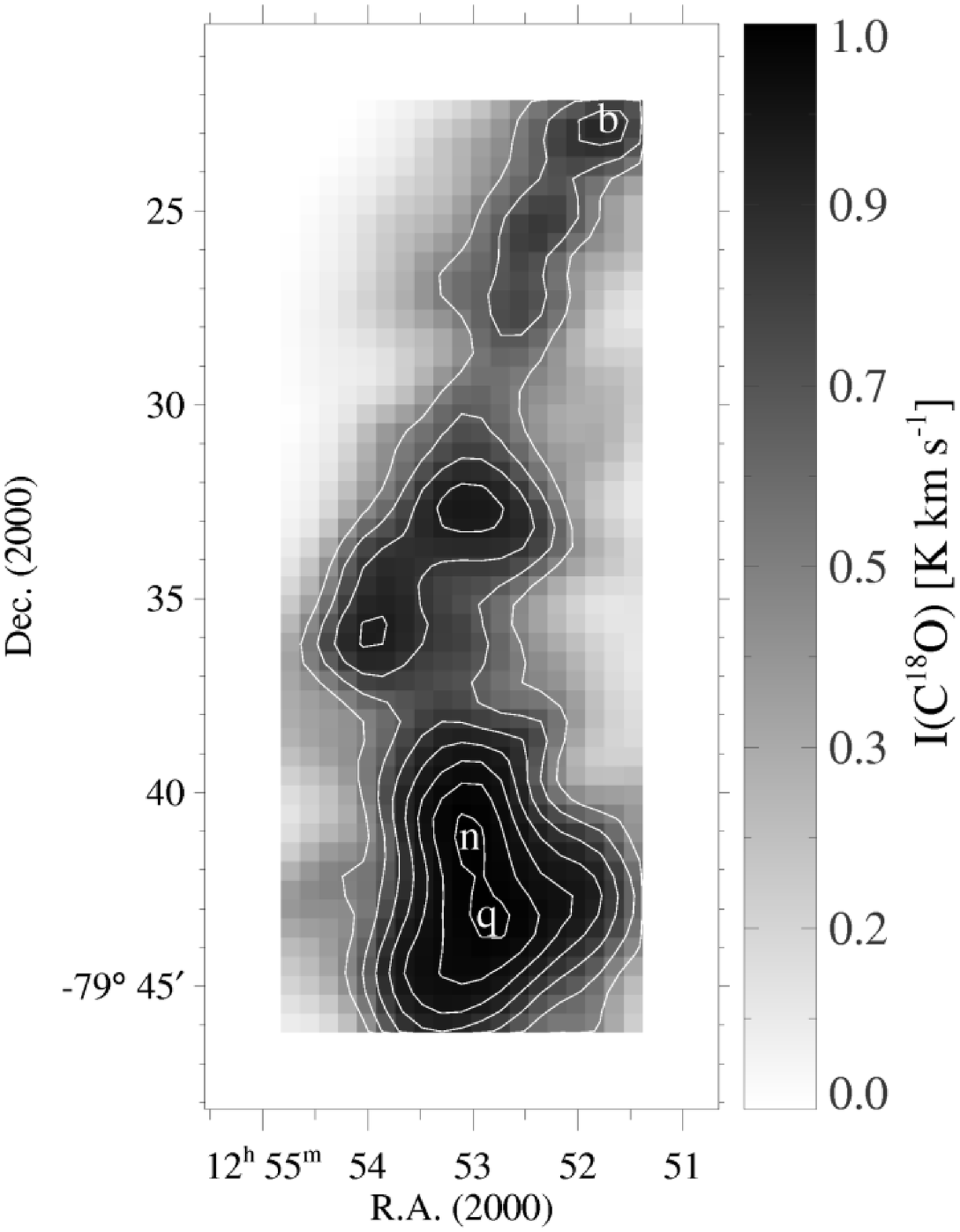}
      \caption{\emph{Left: }Integrated C$^{18}$O intensity map of Chamaeleon I. The map is smoothed to correspond the 2.0$'$ FWHM resolution of the extinction map (Fig. \ref{fig_extinction-map-chaI}). The contour levels start from 0.5 K km s$^{-1}$ and the step is 0.2 K km s$^{-1}$. The numbers give the positions of some CO clumps identified by H05. \emph{Right: }The same for Cha III-B. The contour levels start from 0.4 K km s$^{-1}$ and the step is 0.1 K km s$^{-1}$. The letters mark the most intense clumps identified by Gahm et al. (\cite{gahm02}).
              }
         \label{fig_co-map-chaI}
   \end{figure*}

 The maps of integrated line area of Cha I and Cha III-B are shown in Fig. \ref{fig_co-map-chaI}. Throughout this paper, the C$^{18}$O emission line data are presented on the radiation temperature, $T_\mathrm{R}$, scale. To make the transformation the antenna temperatures,  $T_\mathrm{A}^*$, were divided by the source-beam coupling efficiency $\eta=0.8$, which is in the midway of the main beam efficiency and the moon efficiency of the SEST at this frequency. The intensity map of Cha I reveals extended features, standing out both spatially and in velocity. H05 used the automatic routine 'Clumpfind' of Williams et al. (\cite{williams94}) to find out how these extended features decompose into numerous clumps, and studied their properties in more detail. The clumps with the highest peak intensities were found in the northern arm, close to the Cederblad 112 reflection nebula, with masses of approximately 5 M$_\odot$. High intensity clumps were also found in the central region, and in the southern (and southwestern) parts where the C$^{18}$O emission shows a large local maximum. Some of the most intense clumps are marked with numbers in the Fig. \ref{fig_co-map-chaI}.


The C$^{18}$O column densities along the sightlines were calculated from the integrated intensities using the approximation of optically thin line. The excitation temperature in the clouds was assumed to be 10 K, as was assumed by H05 and Gahm et al. (\cite{gahm02}). The resulting C$^{18}$O column densities reach the maximum values $2\times10^{15}$ cm$^{-2}$ and $2\times10^{14}$ cm$^{-2}$ in Cha I and Cha III-B, respectively.


\subsection{NIR data}
\label{sec_nir-data}

We derive the extinction maps of Cha I and Cha III-B using JHK near-infrared photometry of 2MASS (2 Micron All-Sky Survey). The 10-$\sigma$ limitting magnitudes of the 2MASS data are 15.8, 15.1 and 14.2 in J, H and K bands, respectively. In particular, we use the NICER color excess method (Near-Infrared Color Excess Revised) presented by Lombardi \& Alves (\cite{lombardi01}), which generalizes the color excess method formulated in Lada et al. (\cite{lada94}) for multi-band observations (see Lombardi \& Alves \cite{lombardi01} for the complete description of the method). The data of the 2MASS survey in three bands yield two independent colors, namely $J-H$ and $H-K$, which are used in the following to implement the NICER method. Before any other calculations, the 2MASS photometry is tranformed to the CIT (Caltech Institute of Technology) photometry using the transformation factors given in Eqs. (12)-(15) of Carpenter (\cite{carpenter01}). The color excesses of background stars are scaled to correspond to the extinction in the V band using the diffuse dust reddening law ($R_V=3.1$) of Mathis (\cite{mathis90}):
\begin{equation}
A_\mathrm{V} = 15.87 \times E_{H-K}
\end{equation}
\begin{equation}
A_\mathrm{V} = 9.35 \times E_{J-H}
\end{equation}
It should be noted that the calculated ``visual extinction'' actually is a scaled version of the infrared extinction and does not reflect variations due to possibly altering value of $R_V$.

The 2MASS photometry of two on-source frames and two reference frames was obtained from the 2MASS All-Sky Point Source Catalog \footnote{http://irsa.ipac.caltech.edu/applications/Gator/}. The two on-source fields covered the cloud areas of interest, Cha I and Cha III-B, and they contained approximately 13000 and 3000 stars, respectively. The complete lists of sources in these fields are referred to as the \emph{full samples} in the following. Two reference fields were needed to calculate the intrinsic colors of stars that are free from extinction. The reference fields were selected using the IRAS 100 $\mu$m emission map of the complex as a guideline, and circular areas with radii of 10 arcminutes corresponding to the minima of 100 $\mu$m emission were chosen. The locations of these four fields are outlined in Fig. \ref{IRAS100}.

\subsubsection{Contribution of young stars}

About 200 young stellar objects have been identified within the boundaries of Cha I in previous studies. Due to the surrounding dust envelope these stars generally show NIR-excess emission with respect to the main sequence stars, and thus the extinction value derived towards them is incorrect. Especially in the densest regions, which are likely to contain YSOs, the extinction estimation relies on the photometry of a relatively low number of stars. If one (or several) of these stars possess abnormal colors, the calculated extinction may differ greatly from the ``real'' value. It is therefore important, that the young stars are removed from the sample, even though their number is small compared to the full sample. 

To identify the young stars in Cha I we used the list of known YSOs from Carpenter (\cite{carpenter02}) complemented with the sources from G\'omez \& Kenyon (\cite{gomez01}). The 2MASS sources with positions and colors matching with known YSOs were removed from the sample. Altogether $\sim 90$ \% of the young stars were recovered and removed from the full sample (positional matching). The remaining set of stars, from which the YSOs have been removed, is referred to as the \emph{restricted} sample in the following.

\begin{figure*}[p!]
   \centering
   \includegraphics[width=0.99\columnwidth]{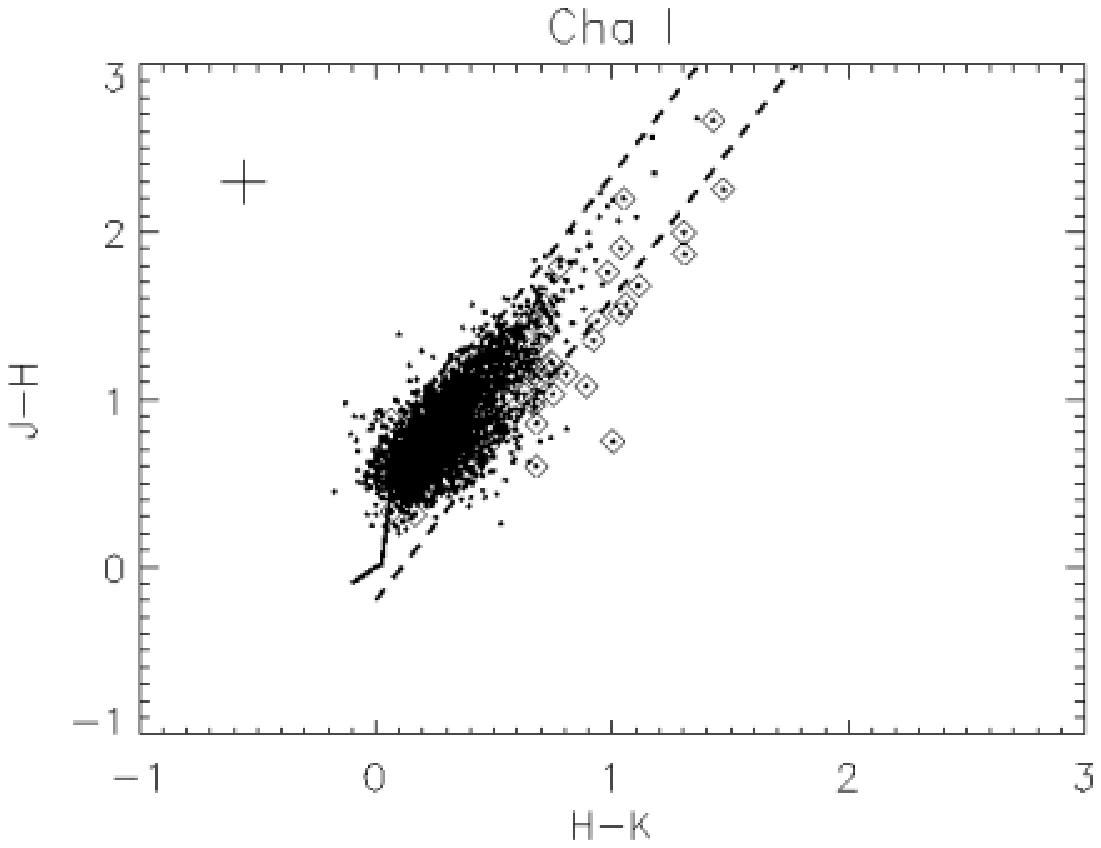} \includegraphics[width=0.99\columnwidth]{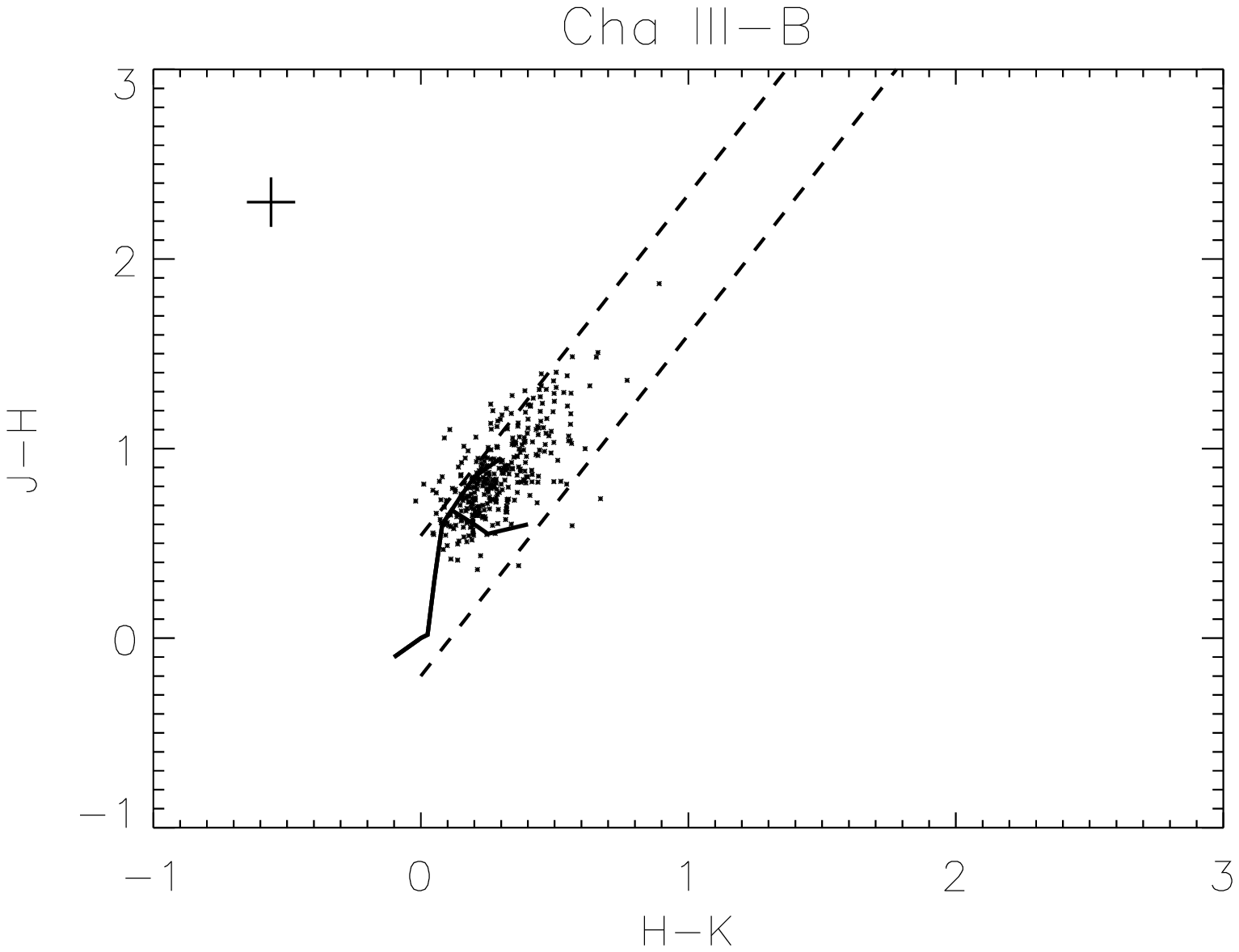}
      \caption{\emph{Left: }Color-Color diagram of 2MASS stars with $S/N > 10$ in Cha I. Diamonds mark the previously identified YSOs from the list of Carpenter (2002). The solid curves are the intrinsic colors of unreddened stars (after Bessell \& Brett 1988), and dashed lines tangenting these curves define the reddening sector. \emph{Right: }The same for Cha III-B. Plus signs in the upper left corner of diagrams are the typical errors of colors.}
         \label{fig_colorcolor-diagrams}100
   \end{figure*}

The color-color diagrams of 2MASS sources having photometry with $S/N > 10$ are presented in Fig. \ref{fig_colorcolor-diagrams}, with intrinsic color-curves of unreddened stars superposed (Bessell \& Brett 1988). The reddening sector is indicated with dashed lines, and it corresponds to the reddening law $E_\mathrm{J-H}/E_\mathrm{H-K}=1.8 \pm 0.03$ derived in Cha I by G\'omez \& Kenyon (\cite{gomez01}). The previously known YSOs are marked in the diagram with diamonds, and obviously no new candidates with high excess emission can be claimed. A few sources fall to the right from the reddening sector around $(H-K, J-H) \approx (0.8, 0.7)$. It is possible that these sources are young stellar objects, even though the colors imply small extinctions and thus small column densities of surrounding dust. Their locations are given in Tab. \ref{tab_new-ysos}.

\begin{table}
\caption{New sources with NIR excess emission in Cha I}             
\label{tab_new-ysos}      
\centering                          
\begin{tabular}{l c c c c}
\hline\hline                 
Star & $\alpha$ (2000.0) & $\delta$ (2000.0) & $J - H$ & $H - K$  \\    
\hline
   1 & 11h 04m 50.8s & -76d 51m 20s & 0.77 & 0.74 \\              
   2 & 11h 04m 59.6s & -77d 35m 58s & 0.82 & 0.81 \\      
   3 & 11h 07m 19.3s & -76d 54m 19s & 0.63 & 0.65 \\
   4 & 11h 10m 57.5s & -77d 44m 44s & 0.75 & 0.69 \\
\hline                                   
\end{tabular}
\end{table}


\subsubsection{Visual extinctions}
\label{sec_visual-extinctions}

\begin{figure*}[p!]
   \centering 
   \includegraphics[width=1.85\columnwidth]{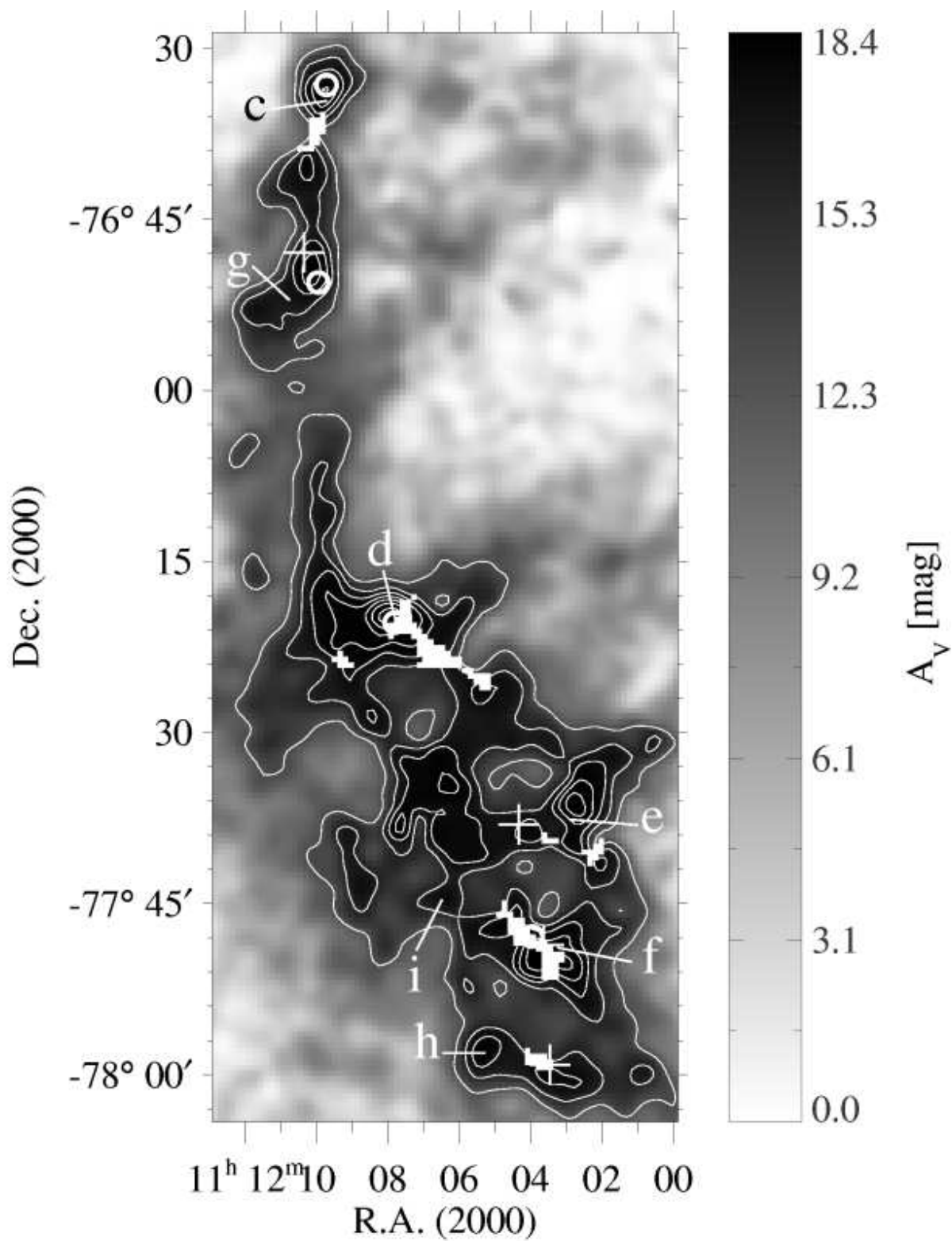}
      \caption{Visual extinction map of Chamaeleon I derived using the NICER color excess method (Lombardi \& Alves 2001) and JHK photometry of 2MASS survey. The FWHM resolution of the map is 2.0$'$. The white pixels result from not having a single star within the FWHM range from the pixel centre, which means that the extinction in these pixels is higher than can be determined. The contour levels start from $A_V = 5^m$ and the step is $2^m$. The letters mark the position of CO cores investigated in Fig. \ref{fig_correlations}. White circles indicate the C$^{18}$O cores of Mizuno et al. (\cite{mizuno99}) and the plus signs the very cold cores detected in the ISOPHOT Serendipity survey by T\'oth et al. (\cite{toth00}).} 
         \label{fig_extinction-map-chaI}
   \end{figure*}

\begin{figure*}
   \centering
   \includegraphics[width=1.5\columnwidth]{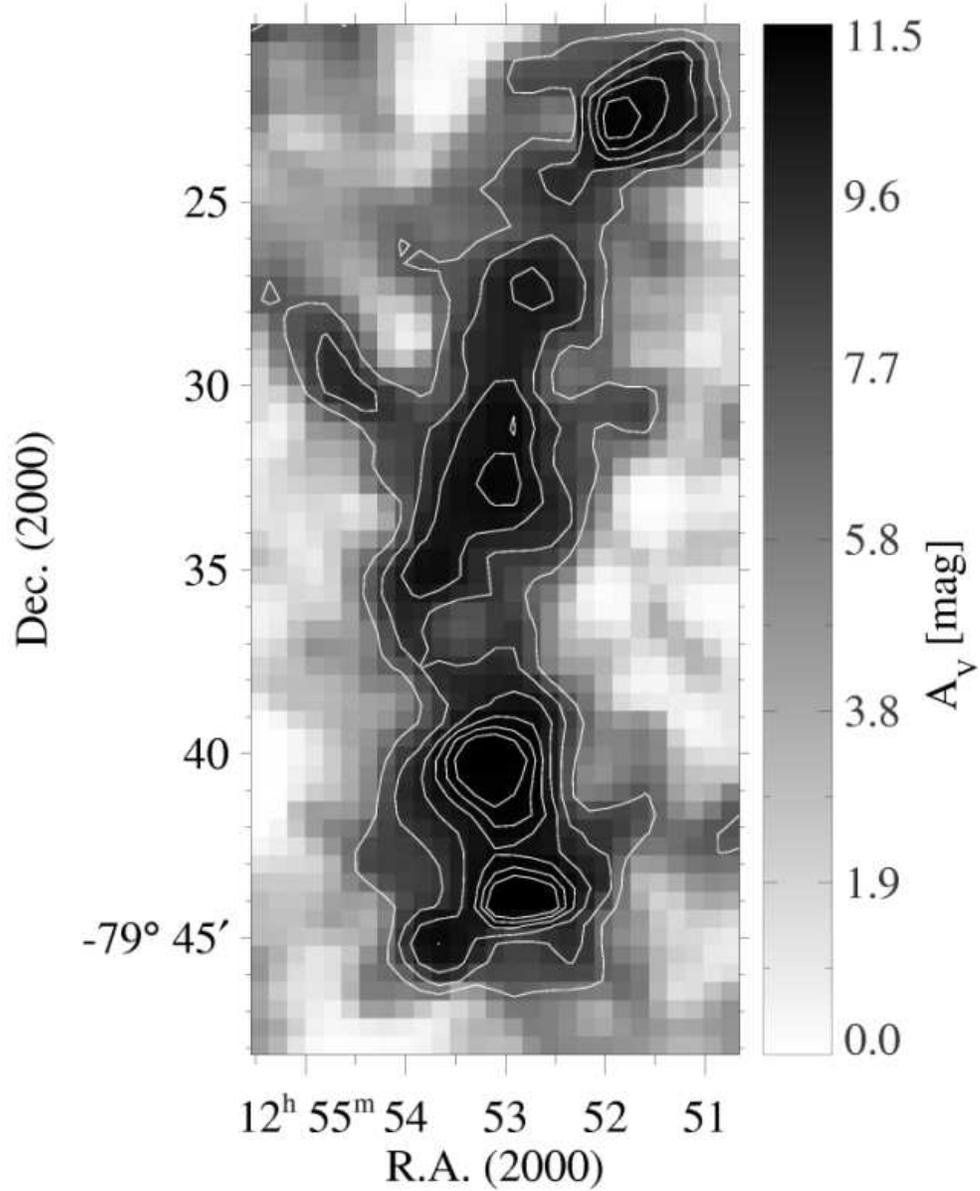}
      \caption{Visual extinction map of Chamaeleon III-B derived using the NICER color excess method (Lombardi \& Alves 2001) and JHK photometry of 2MASS survey. The FWHM resolution of the map is 1.5$'$. The contour levels start from $A_V = 4^m$ and the step is $1^m$.}
         \label{fig_extinction-map-chaIIIb}
   \end{figure*}

Using the restricted sample, from which the contribution of young stars has been eliminated, the NICER color excess method is applied. The extinction in the densest regions of Cha I rises high enough to result in regions of a few arcminutes in size where no 2MASS sources are detected. By requiring at least one star within the FWHM range from the center of each pixel, and estimating the surface density of stars from the reference field, we expect that the extinction estimation is limited to $A_\mathrm{V} \lesssim 20^m$ on the average. This, of course, varies according to what is the true stellar surface density along each sightline. To achieve reasonable spatial resolution over most of the Cha I we select the FWHM resolution of the NICER method to be 2.0$'$, even though it slightly falls below the diameter of some starless regions. The requirement of having at least one star within the FWHM range is thus not met for some pixels, and the determination of extinction in those pixels is impossible. In Cha III-B where the column density is generally smaller we select the FWHM resolution to be 1.5$'$. The resulting extinction maps are presented in Figs. \ref{fig_extinction-map-chaI} and \ref{fig_extinction-map-chaIIIb}. 

\begin{figure*}
   \centering
   \includegraphics[width=0.95\columnwidth]{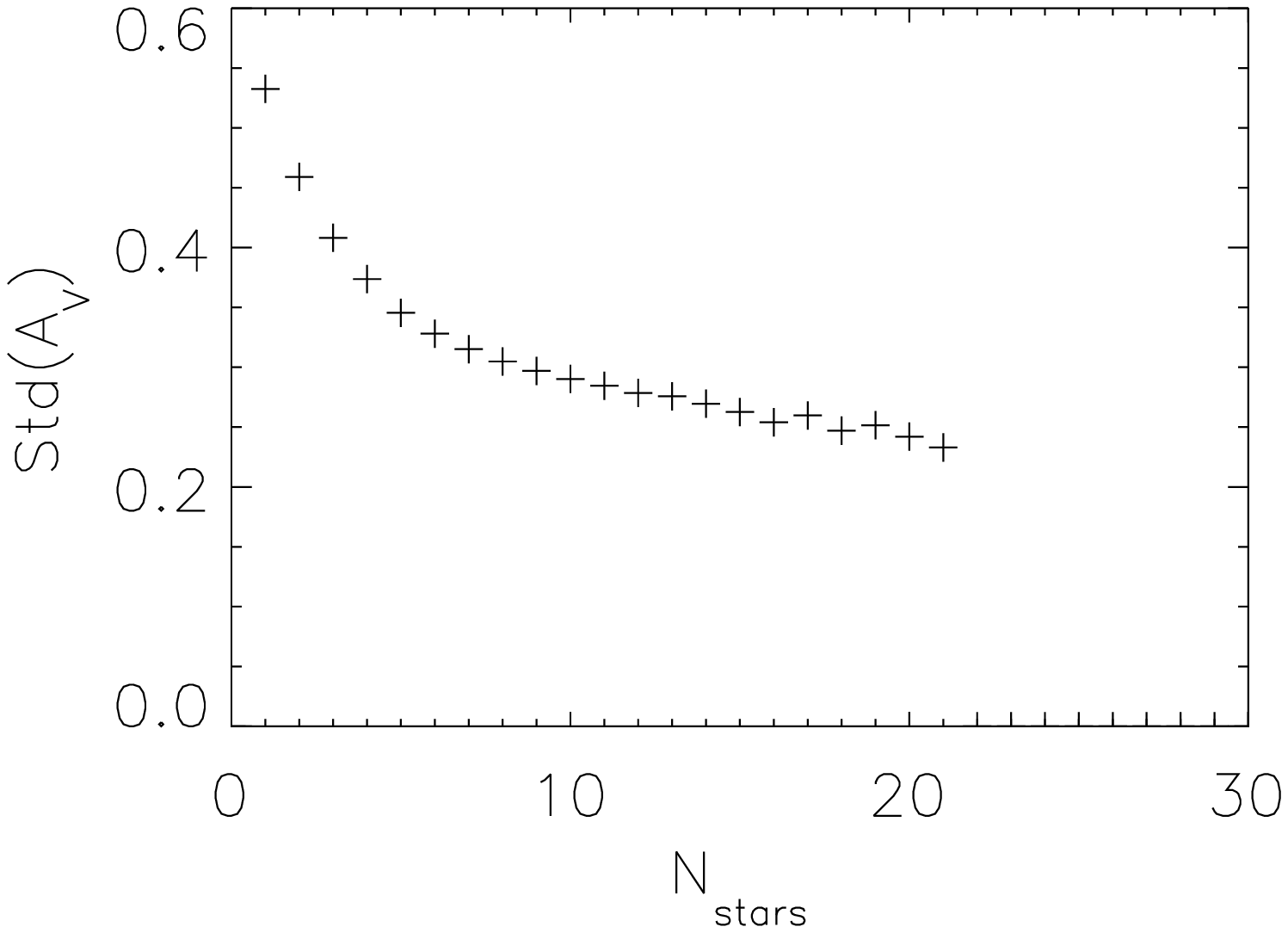} \includegraphics[width=0.95\columnwidth]{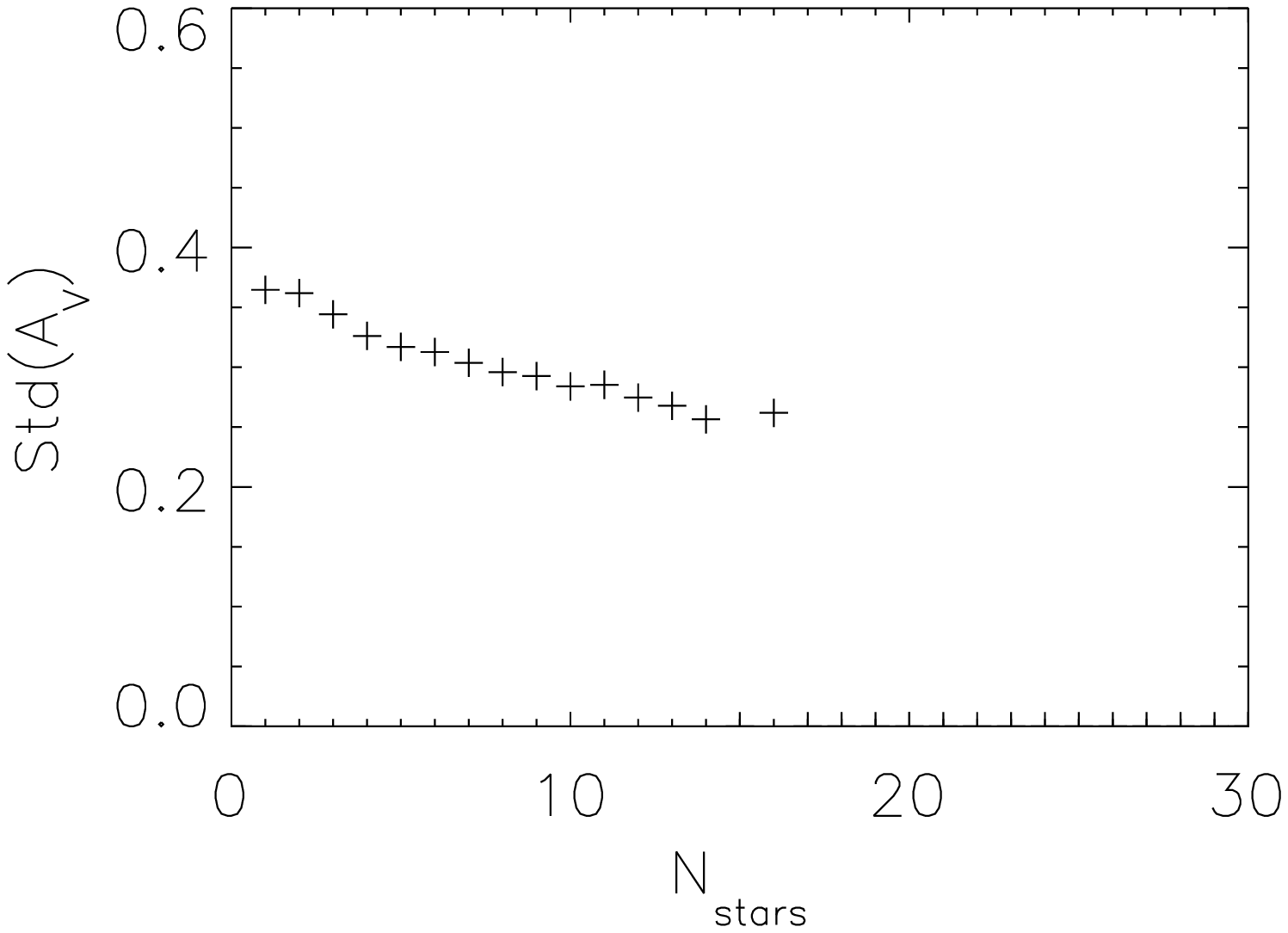}
      \caption{\emph{Left: }The average standard deviation of extinction as a function of star density in Cha I. The x-axis gives the number of stars within the FWHM range from the pixel center. \emph{Right: }The same for Cha III-B.}
         \label{fig_variance-plot}
   \end{figure*}

The errors of the derived extinction maps originate from the uncertainty of the intrinsic colors, $(J-H)_0$ and $(H-K)_0$, and the uncertainty of the observed magnitudes of the on-field stars (Eqs. 9 and 16 in Lombardi \& Alves \cite{lombardi01}). Generally, the 1-$\sigma$ relative error over the maps is about 6\%. On average, the error of extinction per pixel is expected to increase with decreasing star density in denser regions, as the photometric errors become larger. Fig. \ref{fig_variance-plot} describes the behaviour of the average standard deviation of pixels as a function of the number of stars within the FWHM radius. In the low column density regions there are over twenty stars within the FWHM range and the average standard deviation is about 0.2$^m$. When the star density drops, the average standard deviation increases and reaches about 0.5$^m$ at point when there is only one star within the FWHM range. The maximum error in this case is about 1.5$^m$, which equals to the relative error of about 8 \%. Thus, even in the case where the extinction measurement relies on a small number of stars, we find the error acceptable. In Cha III-B the extinction is generally lower and the probability of having apparently brighter stars within the FWHM range is larger. This means that the average photometric errors are lower than in the region of higher extinction. Thus, the increase of the average error at low star densities is not as steep as in Cha I.

The extinction maximum in Cha I is located in the Cederblad 110 region (central part), and it rises up to 19$^m$ at the selected resolution. As mentioned above, extinction in some pixels remains undefined as no stars lie within the radius defined by the FWHM resolution (these ``empty'' pixels are marked white in the extinction map). The extinction in these regions is higher than can be determined. In the Cederblad 112 region (northern part) the maximum extinction is about 15$^m$, and the ``arm'' southwards from Cederblad 112 has $A_\mathrm{V} \gtrsim 10^m$. Numerous clumps with $A_\mathrm{V} > 10^m$ can be found all over the cloud, even in the more quiescent southern part. A large number of less opaque clumps with $A_\mathrm{V} \approx 5^m$ are present on the outskirts of the cloud, being most clearly visible west of the northern arm.

All four C$^{18}$O cores detected by Mizuno et al. (\cite{mizuno99}) located within the borders of the extinction map are clearly seen as local extinction maxima. The southernmost of these appears to have very high column density as the 2MASS source density decreases to zero. In addition, most of the very cold cores (VCCs) detected in the ISOPHOT Serendipity Survey by T\'oth et al. (\cite{toth00}) in this region are recovered as well. See Fig. \ref{fig_extinction-map-chaI} for the positions of these cores.

In H05 the central and southern regions of Cha I are revealed as concentrations of several clumps, some of which remain unresolved at the resolution of the extinction map. Even though the clumps are not individually resolved, their integrated effect is visible. The most obvious concentrations detected are clumps 1 and 2 in the north, 4 and 7 in the east, 5 and 6 in the south and 8 (among others) in the center. Also many less intense clumps, such as 10, 12 and 21, are particularly well resolved. As a bottom line, even most of the low intensity (and less massive) C$^{18}$O clumps such as 35, 38 and 53 seem to have a well defined counterparts in the $A_\mathrm{V}$ map.

Cha III-B has the maximum extinction of about 10$^m$ occuring in the southern part of the studied area. The region of the highest extinction breaks into two maxima, which is also seen (more clearly due to better resolution) in the Fig. 7 of Gahm et al. (\cite{gahm02}). Many smaller features of the C$^{18}$O map such as the local maximum in the north and the short arm leading southeast from the southern maximum are also visible, but somewhat smoothed by the lower resolution of the extinction map.


\section{Correlations between $N$(C$^{18}$O) and A$_\mathrm{V}$}
\label{sec_correlations}

After smoothing the $A_\mathrm{V}$ and $N$(C$^{18}$O) maps of Cha I and Cha III-B to the common resolution of 2.0$'$, the C$^{18}$O column densities and visual extinctions are compared on a pixel-by-pixel basis. The resulting correlation diagrams for Cha I and Cha III-B are are shown in Figs. \ref{fig_correlations}a and b. A linear fit to all data points in Cha I gives the following relationship:\begin{eqnarray}
\lefteqn{ N(\mathrm{C}^{18}\mathrm{O}) = (1.9 \pm 0.1)\times10^{14} A_\mathrm{V} {} } \nonumber\\
& & {} \qquad \qquad - (4.9 \pm 0.7)\times10^{14} \qquad \mathrm{(Cha I)}
\label{eq_result-in-chaI}
\end{eqnarray}
The plot in Fig. \ref{fig_correlations}a suggests that the proportionality is steeper than the fit at low values of $A_\mathrm{V}$, but becomes slightly shallower above $A_\mathrm{V} \gtrsim 7^m$, and finally at $A_\mathrm{V} \gtrsim 12^m$ most datapoints lie below the fitted line.

In Cha III-B we obtain the following fit:
\begin{eqnarray}
\lefteqn{ N(\mathrm{C}^{18}\mathrm{O}) = (2.0 \pm 0.1)\times10^{14} A_\mathrm{V} {} } \nonumber\\
& & {} \qquad \qquad - (3.9 \pm 0.4)\times10^{14} \qquad \mathrm{(Cha IIIB)}
\label{eq_result-in-chaIIIb}
\end{eqnarray}
In Cha III-B the extinction remains below 10$^m$, and no clear deviation from the linear relationship can be seen. The parameters of the linear fits to the Cha I and Cha III-B data sets do not differ significantly at this scale, where the relation is averaged over the Cha I cloud.

The extinction map of Cha I has several opaque spots with $A_\mathrm{V} > 10^m$. Almost all of them can be identified with C$^{18}$O clumps found by H05. These clumps with radii ranging from about 4$'$ to 10$'$ can be discerned also on the C$^{18}$O map smoothed to 2$'$. The $N$(C$^{18}$O) vs. $A_\mathrm{V}$ plots of the regions of seven selected clumps are shown in Figs. \ref{fig_correlations}c-i. Figs \ref{fig_correlations}c and i represent regions where the  $N$(C$^{18}$O)/$A_\mathrm{V}$ ratio is higher, whereas Figs. \ref{fig_correlations}e and h represent regions where the $N$(C$^{18}$O)/$A_\mathrm{V}$ ratio is lower than expected from the linear fit to the whole Cha I.


\section{Discussion}
\label{sec_discussion}

As the CO molecule is chemically very stable, the variations of the
$N({\rm C^{18}O})/A_\mathrm{V}$ ratio depend on the
accretion of molecules onto grain surfaces and on desorption processes
which release them back into the gas phase. In the obscured and dense cores with no central source the abundance drop of CO, caused by accretion onto the surfaces of cold dust grains, may be several orders of magnitudes. The dust temperature is very low, about $\sim$ 10 K and evaporation due to thermal desorption very unefficient. 
The rate of the thermal desorption can be written as:
\begin{equation}
\xi(M) \propto \mathrm{exp} { \Big\{ -\frac{E_\mathrm{b}(M)}{kT_\mathrm{d}} \Big\} } 
\end{equation}
where $E_\mathrm{b}(M)$ is the binding energy, $T_\mathrm{d}$ the dust temperature and $M$ refers to the composition of the grain mantle (Watson \& Salpeter \cite{watson72}). Non-thermal processes such as desorption caused by cosmic rays are also present, and in fact required to sustain the observed abundances in cold cores (Willacy and Williams \cite{willacy93}; Hasagawa and Herbst \cite{hasegawa93}). However, in the presence of a central heating source the thermal desorption of CO quickly overcomes other processes due to the low binding energy of the CO molecule. In the envelope surrounding the heating source the CO is evaporated in a short timescale, and most of the CO should remain in gas phase as the dust temperature keeps rising (e.g. Rodgers \& Charnley \cite{rodgers03}; Lee et al. \cite{lee04}).

The depletion of gas phase C$^{18}$O can be quantified with the aid of
the depletion factor, $f_{\rm D}$, which is defined as the ratio of
the total fractional abundance to the gas phase abundance: $f_{\rm D}
\equiv \chi({\rm C^{18}O,gas+dust})/\chi({\rm C^{18}O,gas})$. As
discussed in Hotzel et al. (\cite{hotzel02}), in an isothermal cloud where these
processes are in equilibrium $f_{\rm D}$ is proportional to the gas
density according to $f_{\rm D}=1+C n_{\rm H_2}$, where $C=A/B$ is the
ratio of the adsorption and desorption coefficients. These
coefficients depend on the grain properties, temperature and the flux
of heating particles (primarily cosmic rays). A reasonable set of
parameters gives the following estimate in dark clouds: $C \approx 5\,
10^{-4} \,{\rm cm}^3$ (Hotzel et al. \cite{hotzel02}). Large changes in adsorption
and desorption coefficients are not likely to occur within a molecular cloud,
unless local heating by radiation or shocks increases the grain
temperature above a critical temperature of CO desorption, which lies
in the range 20-30 K (Collings et al. \cite{collings03}). The no-depletion fractional C$^{18}$O abundance, $\chi({\mathrm C^{18}O,gas+dust})$, is
determined by the gas-phase production of CO in the low-density cloud
envelope, and is probably constant throughout the cloud.

The density dependence of the depletion factor is transferred also 
to the observed $N({\rm C^{18}O})/A_{\rm V}$ ratio:
\begin{eqnarray}
\frac{N({\rm C^{18}O, gas})}{A_{\rm V}} &=& 
\frac{\int n({\rm C^{18}O, gas}) \, ds} {\alpha \int n({\rm H_2}) \, ds} 
\nonumber \\
&=& \frac{\int n({\rm C^{18}O,gas+dust})/(1+C  n({\rm H_2}))ds}
                         {\alpha \int n({\rm H_2}) \, ds} 
\nonumber \\
&=& \frac{\int \chi({\rm C^{18}O,gas+dust}) \frac{n({\rm H_2})}{1+C  n({\rm H_2})} 
    \, ds}  {\alpha \int n({\rm H_2}) \, ds} 
\end{eqnarray} 
The $A_{\rm V}$ to $N({\rm H_2})$ conversion factor, $\alpha$, is given in
Eq. \ref{eq_bohlin2}. For example, assuming line-of-sight homogeneity one obtains the following 
relation:
\begin{equation}
\frac{N({\rm C^{18}O, gas})}{A_{\rm V}} = 
\frac{\chi({\rm C^{18}O,gas+dust})}{\alpha} \frac{1}{1+C\, n({\rm H_2})} \; .
\end{equation} 
Although this assumption is not very realistic, the example
illustrates the tendency that the measured $N({\rm C^{18}O})/{A_{\rm
V}}$ ratios are likely to be smaller along line of sights which 
contain very dense gas than elsewhere.

Based on the arguments above, the variations of the $N$(C$^{18}$O) / $A_\mathrm{V}$ ratio can be attributed to two main processes which can be seen as consequencies of gravitational contraction. On the one hand, a large column density required for C$^{18}$O detection shelters C$^{18}$O efficiently from the external radiation field, and as the accretion rate of molecules onto the grain surfaces depends strongly on the density, a high degree of CO depletion can be expected towards dense cores without internal heating sources. On the other hand, the densest cores are likely to be active sites of star formation which can give rise to an increased local heating releasing molecules from grains. Also violent events such as winds and jets from newly born stars may play a role in this process.

Apart from the physical reasons, the validity of the assumptions made in the calculation of CO column densities may vary as a function of density (i.e. optical depth). The comparison of C$^{18}$O and C$^{17}$O line profiles in H05, however, supports the view that the C$^{18}$O line remains optically thin. The assumption of the constant excitation temperature (10 K) is perhaps more fragile in this sense, as the temperatures in cloud cores have been predicted to drop towards the core center (Zucconi et al. \cite{zucconi01}; Galli et al. \cite{galli02}) and observations have also given such indications (e.g. Ward-Thompson et al. \cite{ward-thompson02}; Lehtinen et al. \cite{lehtinen04}). The observed temperature gradients are nevertheless quite small, and even going from, say, $T_\mathrm{ex}=8$ K to $T_\mathrm{ex}=15$ K does not produce a large enough difference to explain the observed N(C$^{18}$O) variations in clumps (see the discussion e.g. in Harjunp\"a\"a and Mattila \cite{harjunpaa96}). Furthermore, the neglection of the temperature gradient should lead to similar variations in clumps with equal maximum extinction. The situation in Figs. \ref{fig_correlations}h-i representing clumps with equal maximum extinctions suggests therefore that temperature gradients cannot be the main reason for the variation in the  $N$(C$^{18}$O) / $A_\mathrm{V}$ ratio.

\subsection{$N$(C$^{18}$O) / $A_\mathrm{V}$ ratio in Cha I and III-B}

The $N$(C$^{18}$O) / $A_\mathrm{V}$ relations derived for Cha I and Cha III-B do not differ significantly from each other on the large scale where the whole Chamaeleon I region is considered (Eqs. \ref{eq_result-in-chaI}, \ref{eq_result-in-chaIIIb}). This is not in agreement with the result derived in Hayakawa et al. (\cite{hayakawa01}), in which a significant difference between the clouds was found. In Hayakawa et al. (\cite{hayakawa01}) the slope derived for Cha I was steeper than that of Cha III by a factor of 1.4. The relationships derived by Hayakawa et al. (\cite{hayakawa01}) are quoted below: 
\begin{eqnarray}
\lefteqn{ N(\mathrm{C}^{18}\mathrm{O}) = (3.1 \pm 0.1)\times10^{14} A_\mathrm{V} {} } \nonumber\\
& & {} \qquad \qquad - (4.4 \pm 0.7)\times10^{14} \qquad \mathrm{(Cha I, Hayakawa)}
\end{eqnarray}
\begin{eqnarray}
\lefteqn{ N(\mathrm{C}^{18}\mathrm{O}) = (2.3 \pm 0.3)\times10^{14} A_\mathrm{V} {} } \nonumber\\
& & {} \qquad \qquad - (0.3 \pm 0.8)\times10^{14} \qquad \mathrm{(Cha III, Hayakawa)}
\end{eqnarray}

A possible reason for this difference in results appears to be discrepancy between the C$^{18}$O intensities observed with the SEST and the NANTEN telescopes: the intensities observed using SEST in Cha I are smaller by a factor of 1.5-2.2 compared to intensities at the positions of cores given in Mizuno et al. (\cite{mizuno99}). On the other hand, the intensities of Mizuno et al. (\cite{mizuno99}) in Cha III-B exceed those of the present study only by a factor of $\sim 1.2$. The comparison was done after smoothing the SEST maps to correspond the 2.7 FWHM beam of NANTEN. However, the number of common datapoints especially in Cha III-B is small, and a proper overall comparison is not possible. But if the difference in the intensities is generally at this level, it is alone sufficient to explain the difference between the results of the present study and Hayakawa et al. (\cite{hayakawa01}).

\subsection{$N$(H$_2$) / $N$(C$^{18}$O) ratio and the cloud masses}

To express the observed $N$(C$^{18}$O) / $A_\mathrm{V}$ relation in terms of hydrogen column density we use the $N$(HI+H$_2$) / $E$(B-V) ratio derived from the observations of the Copernicus satellite by Bohlin et al. (\cite{bohlin78}):
\begin{equation}
\frac{N(H+H_2)}{E(B-V)}=5.8\times10^{21}\mathrm{cm}^{-2} \mathrm{mag}^{-1}
\label{eq_bohlin}
\end{equation}
The sightlines of Bohlin et al. (\cite{bohlin78}) were restricted to low extinctions ($A_\mathrm{V} < 2$), but the result has been verified also for denser clouds (Vuong et al. \cite{vuong03}). Adopting the value $R_\mathrm{V}=A_\mathrm{V}/E_\mathrm{B-V}=3.1$, consistent with the implementation in NICER, we have:
\begin{equation}
\frac{N(\mathrm{H}_2)}{A_\mathrm{V}}=9.4\times10^{20} \mathrm{cm}^{-2} \mathrm{mag}^{-1}
\label{eq_bohlin2}
\end{equation}

We first calculate the masses of the clouds from the extinction maps using Eq. \ref{eq_bohlin2}. The masses are 260 M$_\odot$ and 18 M$_\odot$ for Cha I and Cha III-B, respectively. Then, we apply Eq. \ref{eq_bohlin} to Eqs. \ref{eq_result-in-chaI} and \ref{eq_result-in-chaIIIb}, and have:
\begin{equation}
N(\mathrm{H}_2) = 4.9 \times 10^6 N(\mathrm{C}^{18}\mathrm{O}) + 2.4 \times 10^{21} \mathrm{cm}^{-2} \qquad \mathrm{(Cha I)}
\label{eq_h2-c18o-relation-chaI}
\end{equation}
\begin{equation}
N({\sf H}_2) = 4.7 \times 10^6 N(\mathrm{C}^{18}\mathrm{O}) + 1.8 \times 10^{21} \mathrm{cm}^{-2}\qquad \mathrm{(Cha IIIB)}
\label{eq_h2-c18o-relation-chaIIIb}
\end{equation}
These relations are actually very close to the ``canonical'' C$^{18}$O / $A_\mathrm{V}$ values derived by Frerking et al. (\cite{frerking82}): 
\begin{equation}
N(\mathrm{H}_2) = 5.9 \times 10^6 N(\mathrm{C}^{18}\mathrm{O}) + 1.3 \times 10^{21} \mathrm{cm}^{-2} \qquad \mathrm{(Taurus)}
\end{equation}
\begin{equation}
N(\mathrm{H}_2) = 5.9 \times 10^6 N(\mathrm{C}^{18}\mathrm{O}) + 3.9 \times 10^{21} \mathrm{cm}^{-2} \qquad \mathrm{(}\rho \  \mathrm{Ophiuchi)}
\end{equation}

The masses of the observed areas using Eqs. \ref{eq_h2-c18o-relation-chaI} and \ref{eq_h2-c18o-relation-chaIIIb} are 280 M$_\odot$ and 21 M$_\odot$. The masses derived using the two different approaches are not independent and turn out to be, as expected, close to each other. This reflects the fact that most of the mass comes from the regions with moderate density, that is, from the region where the linear  $N$(C$^{18}$O) / $A_\mathrm{V}$ relation holds well on the average.

\subsection{Variations of $N$(C$^{18}$O) / $A_\mathrm{V}$ in Cha I}

\begin{figure*}[p!]
   \centering
   \includegraphics[width=0.75\columnwidth]{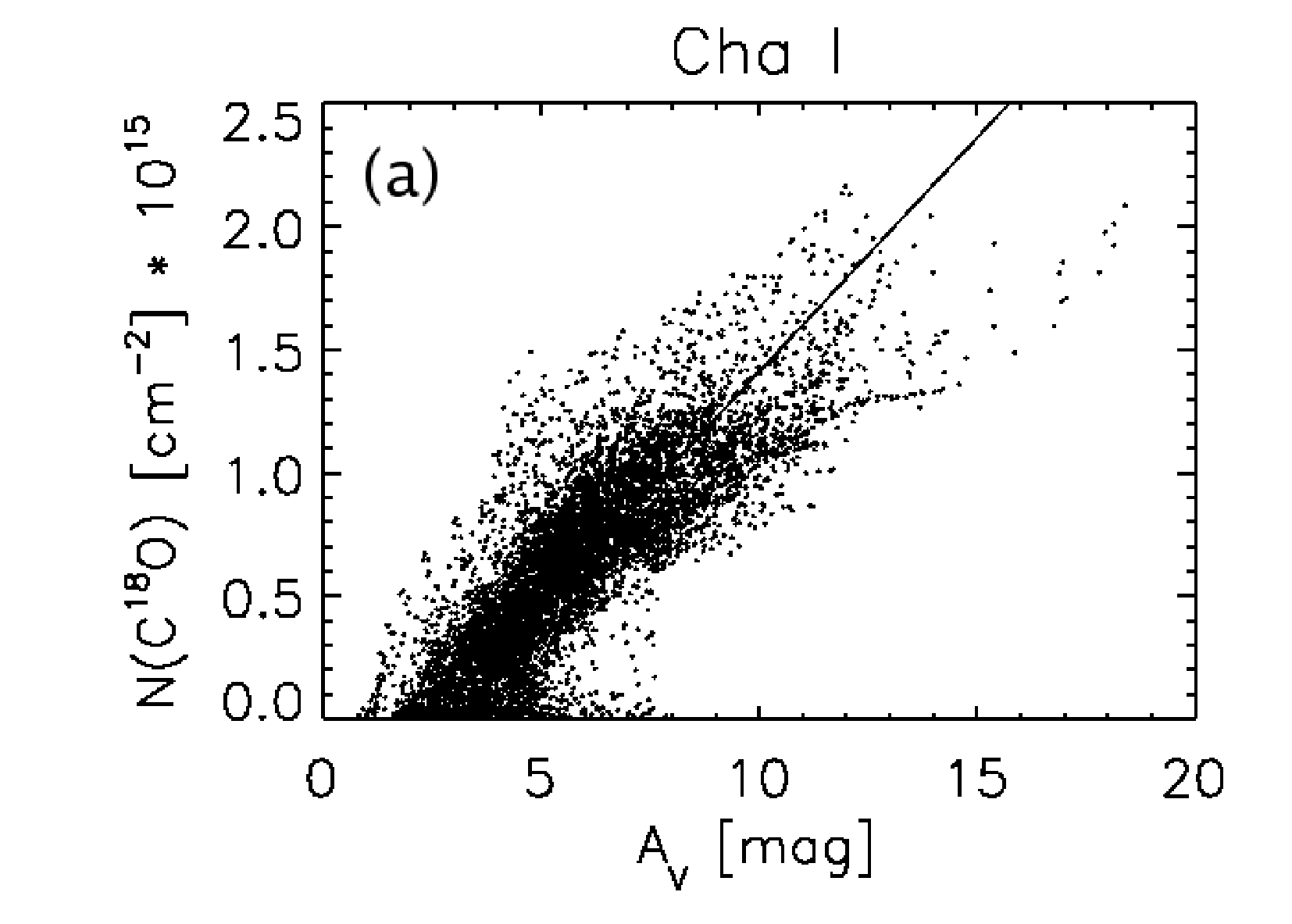} \includegraphics[width=0.75\columnwidth]{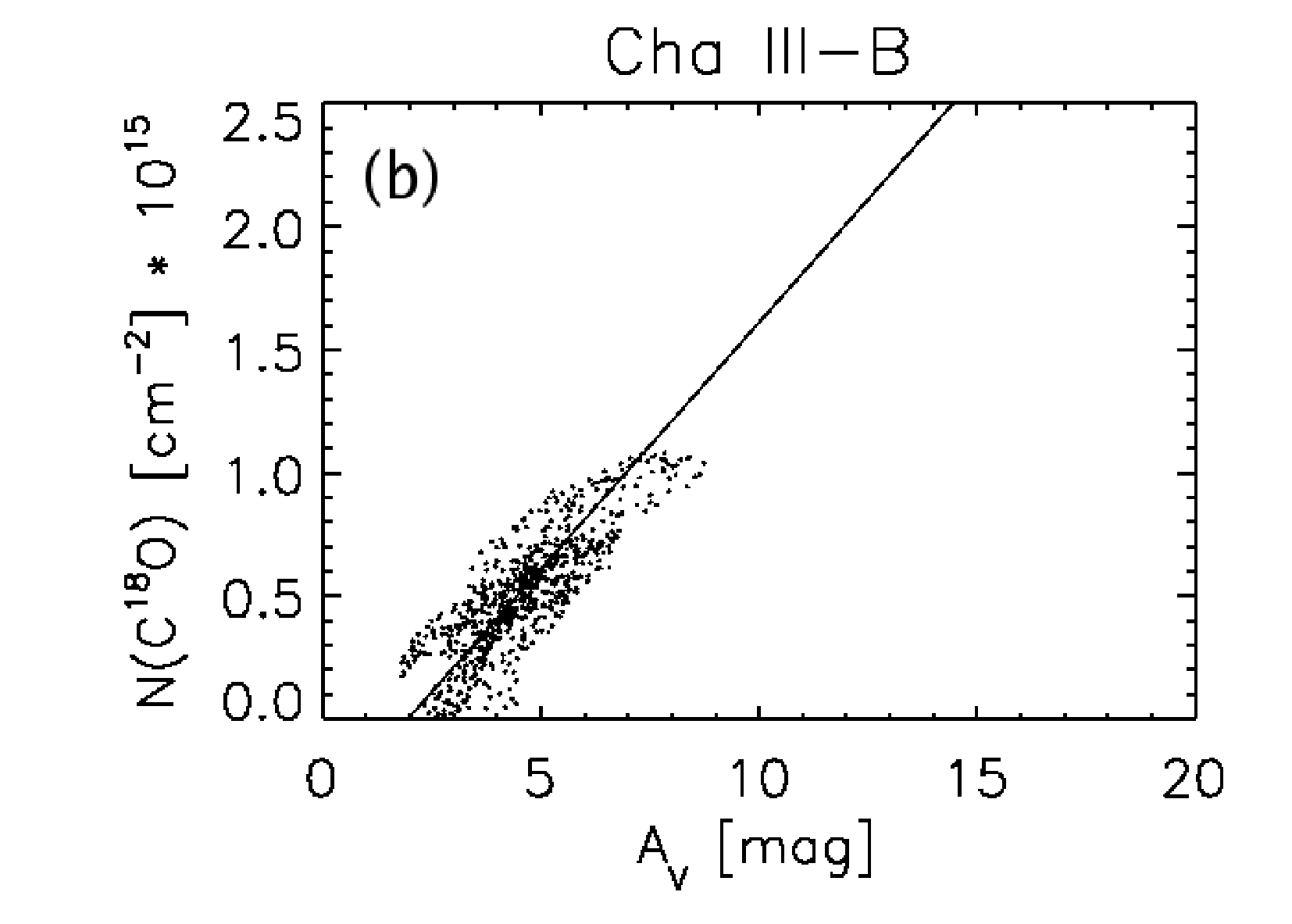} 

\includegraphics[width=0.75\columnwidth]{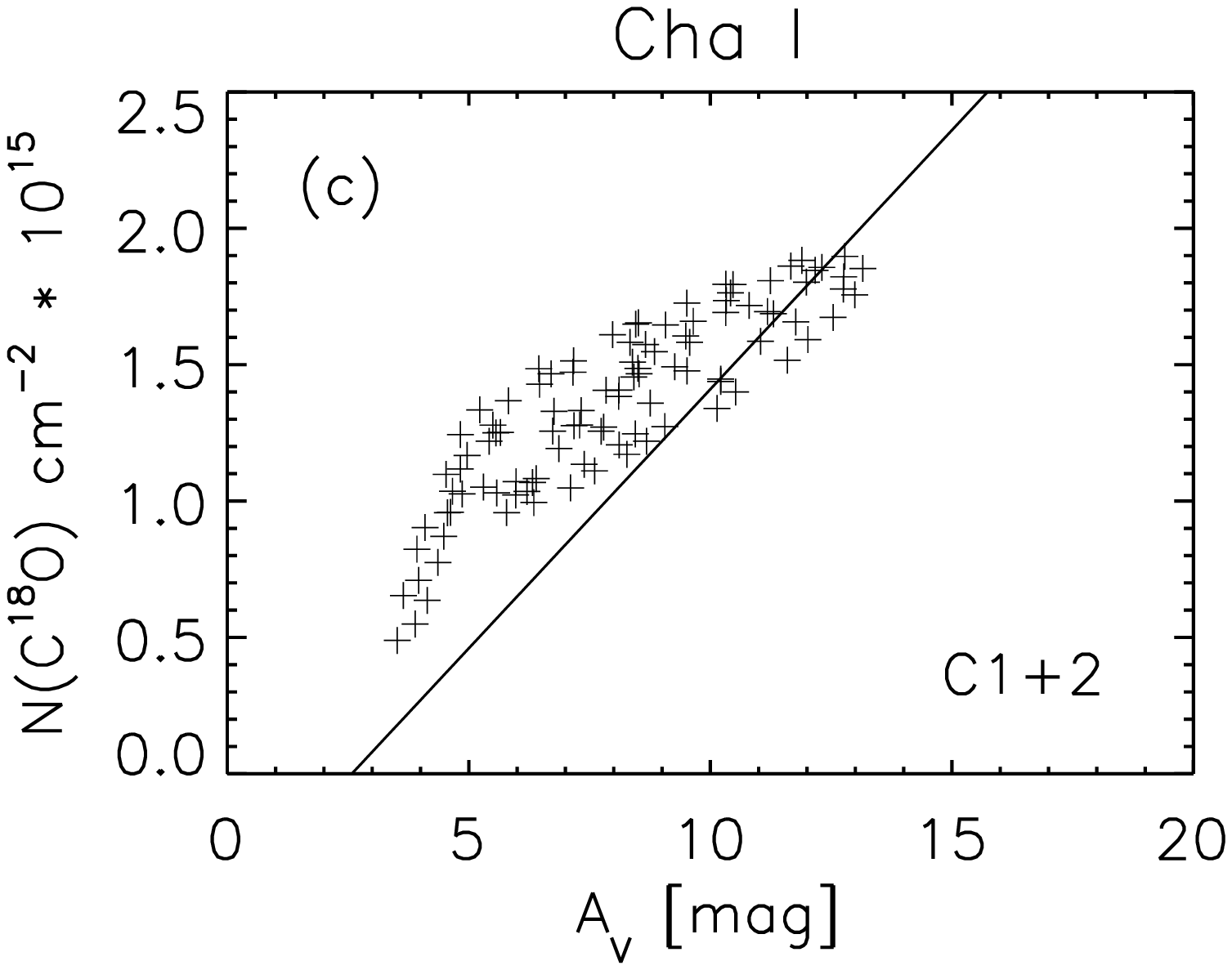}\includegraphics[width=0.75\columnwidth]{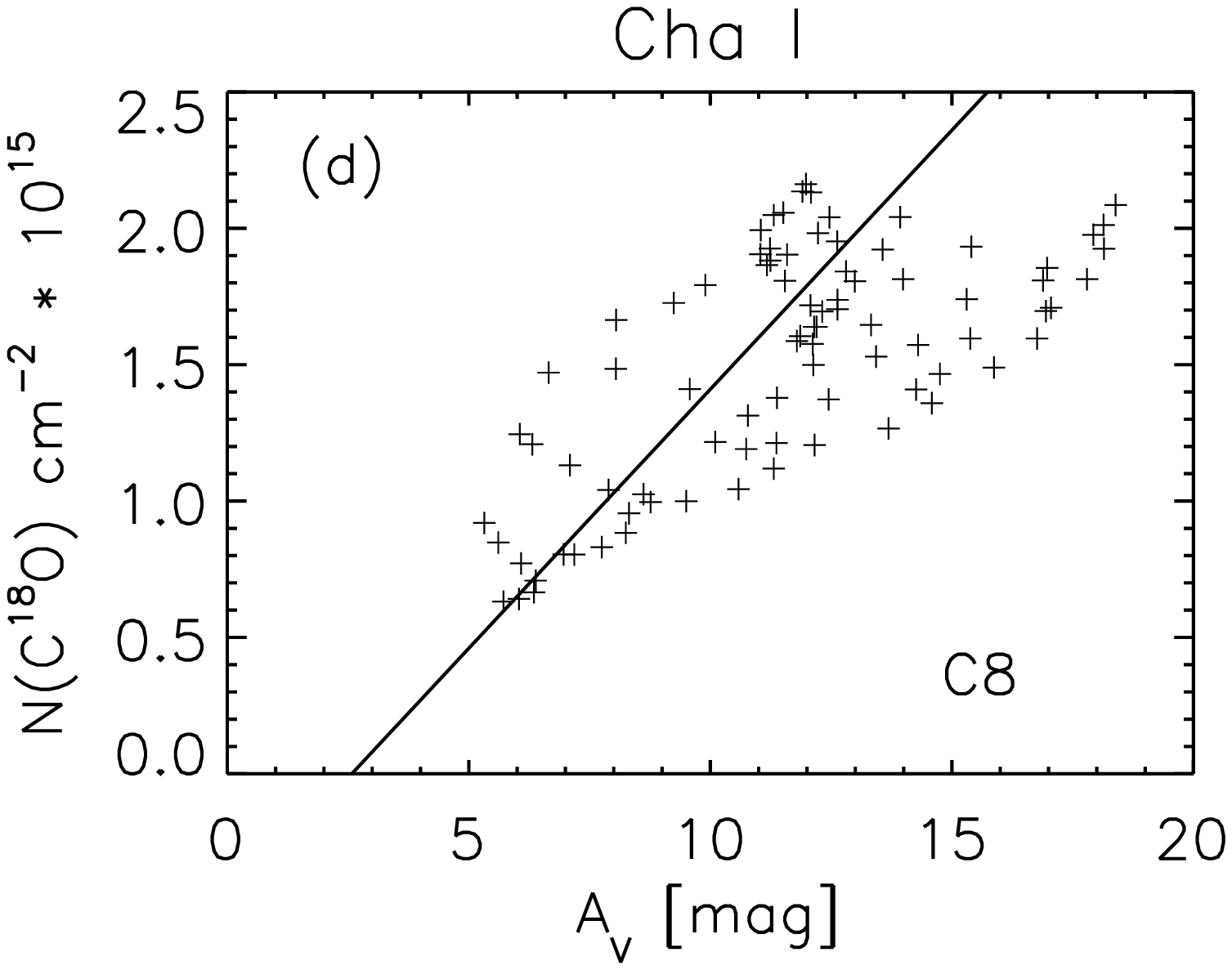} 

\includegraphics[width=0.75\columnwidth]{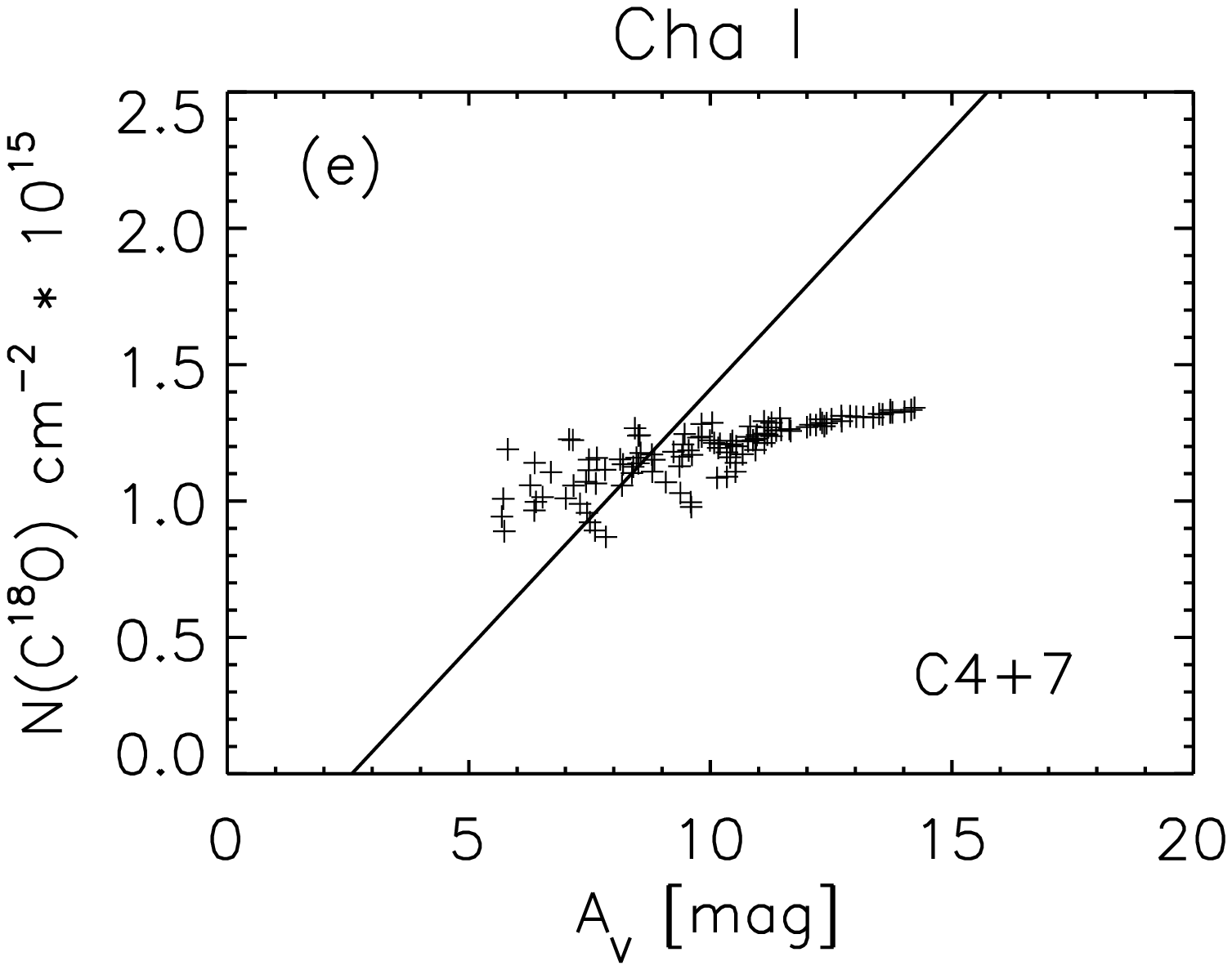} \includegraphics[width=0.75\columnwidth]{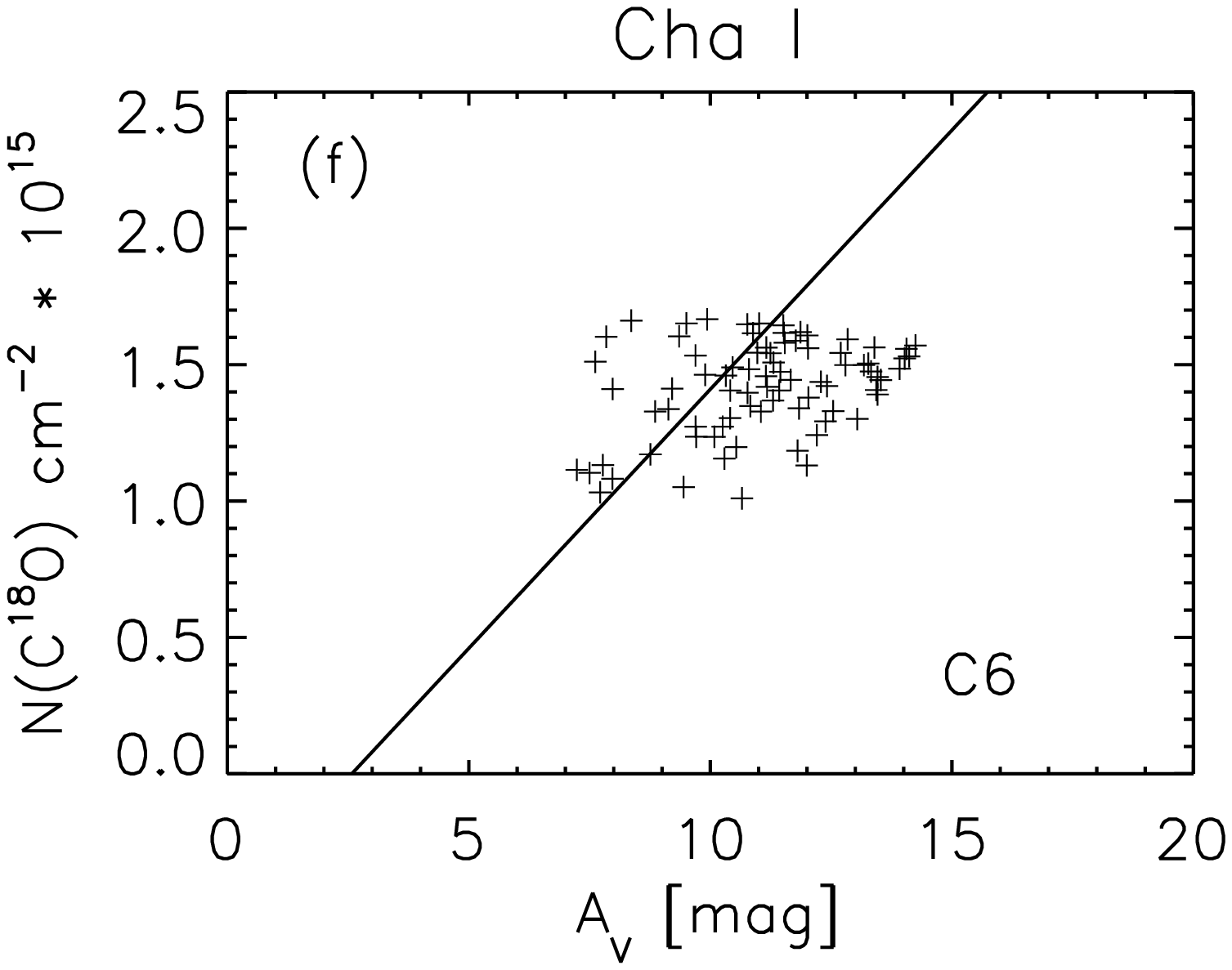}

\includegraphics[width=0.75\columnwidth]{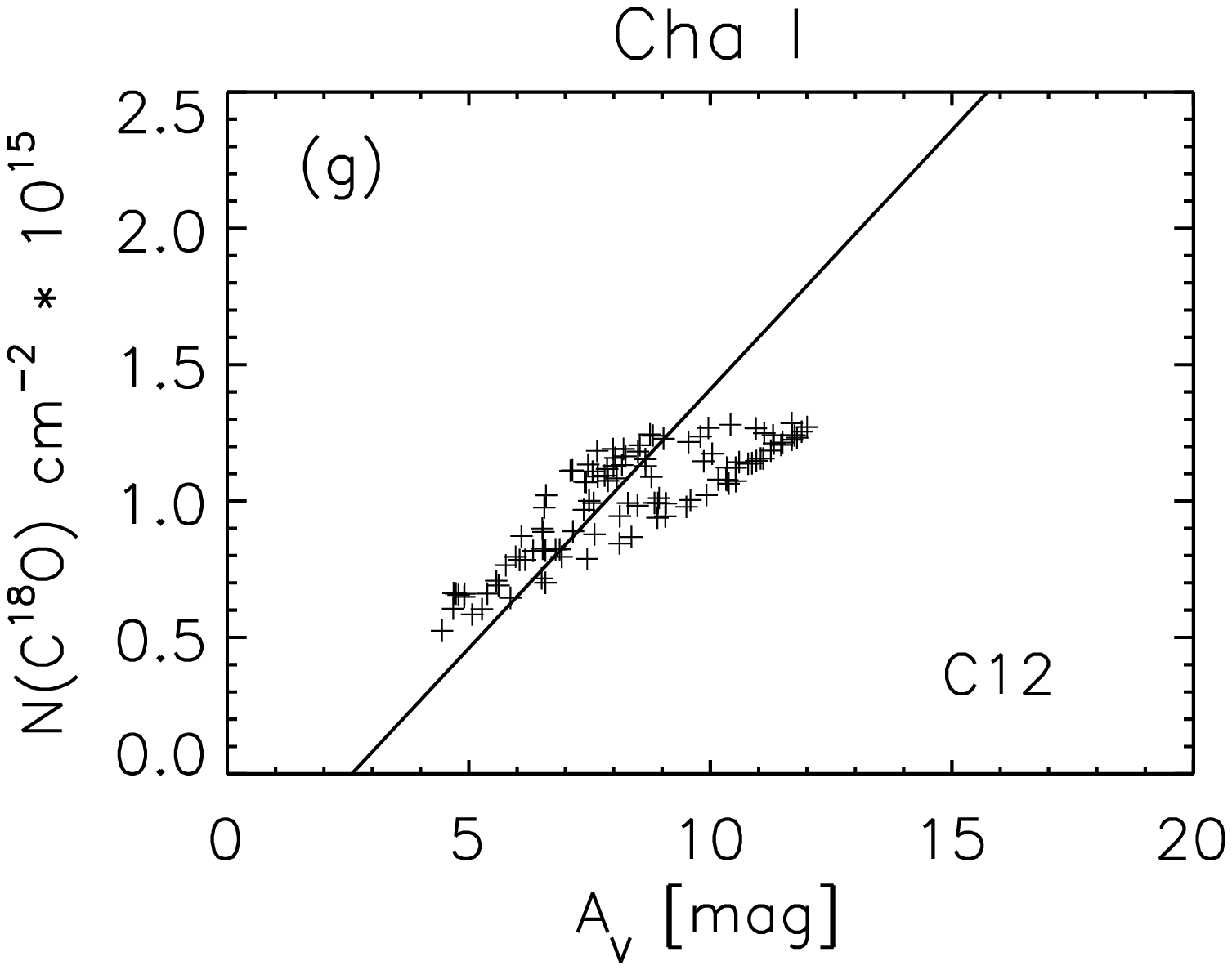} \includegraphics[width=0.75\columnwidth]{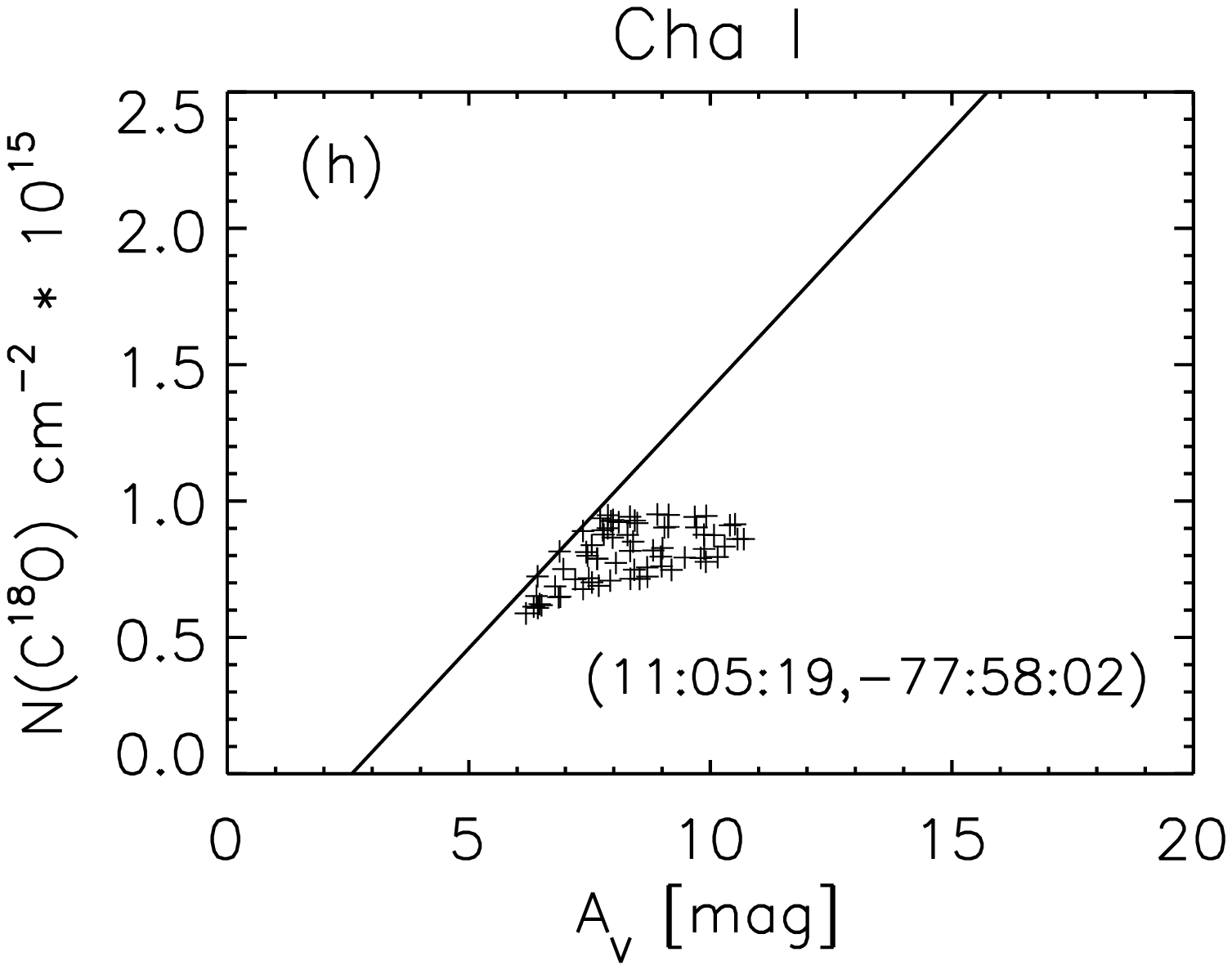}

\includegraphics[width=0.75\columnwidth]{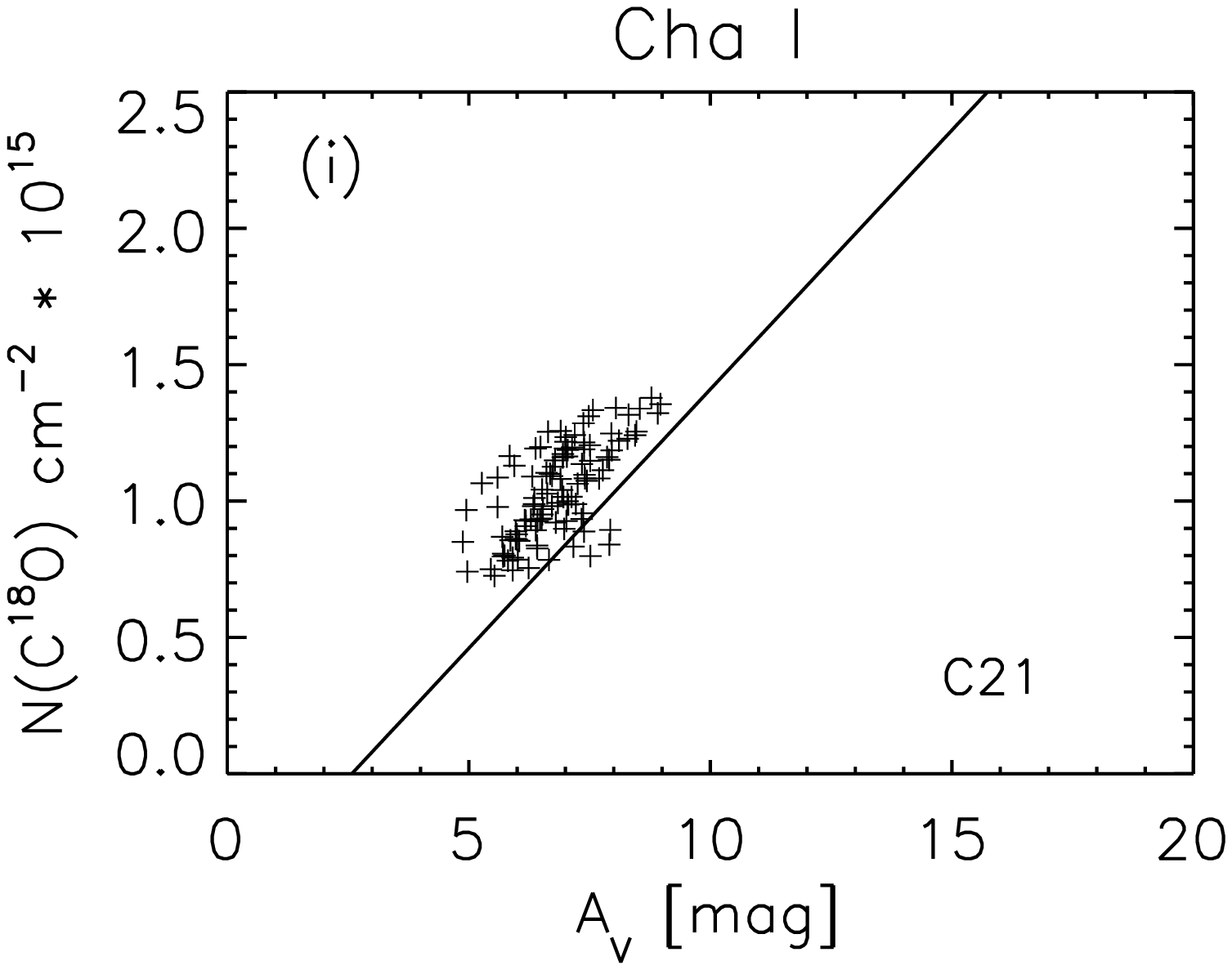}

      \caption{The correlations of $N$(C$^{18}$O) and $A_\mathrm{V}$ in Cha I \emph{(a)}, Cha III-B \emph{(b)}, and in selected clumps of Cha I. The clump locations following Haikala et al. (2005): \emph{(c) }clumps 1 and 2, \emph{(d) }clump 8,\emph{(e) }clumps 4 and 7, \emph{(f) }clump 6, \emph{(g) }clump 12, \emph{(h) }(ra, dec) = (11:05:19, -77:58:02), \emph{(i) }clump 21. The locations of the clumps are shown in Fig. \ref{fig_co-map-chaI}.
              }
         \label{fig_correlations}
   \end{figure*}

Figs. \ref{fig_correlations}c-i point out differences in the $N$(C$^{18}$O) vs. $A_\mathrm{V}$ relation when comparing individual clumps in Cha I. The ratio in the northern region differs significantly from the rest of the cloud, and small differences between other clumps can be seen mainly in the overall C$^{18}$O level (the intercept of the linear model). Also, few interesting exceptions are found, especially in the western and southern parts of the cloud.

The combination of clumps 1 and 2 in the north represents the high-CO side of the $N$(C$^{18}$O) vs. $A_\mathrm{V}$ relation over the full A$_\mathrm{V}$ scale (Fig. \ref{fig_correlations}c). Compared with clumps in less active regions the average $N$(C$^{18}$O) level implies the overabundance of C$^{18}$O by a factor of $1.3 \pm 0.2$ with respect to the relation averaged over the whole Cha I, and by a factor of $3.5 \pm 0.5$ with respect to the relation of the clump with low C$^{18}$O level in Fig. \ref{fig_correlations}h. This supports the view that in this region the C$^{18}$O abundance is indeed enhanced due to very high star formation activity. Along the ``arm'' southwards from clumps 1 and 2 there is another very massive clump, clump 12, which has no identified YSOs in it. The extinction rises close to the value of the clumps 1 and 2, i.e. $\sim 12^m$, but the C$^{18}$O level remains significantly lower. The plot shown in Fig. \ref{fig_correlations}g is different from that of clumps 1 and 2. Again, this tendency could be interpreted in terms of less intense radiation, and thus higher overall level of gas depletion.

The active region in the central part of Cha I has a very high column density with $A_\mathrm{V} > 15^m $ over a large area. For several tens of pixels the star density drops to zero per NICER beam, i.e. the FWHM range of NICER, and the extinction cannot be determined. Also, relying on very few stars per beam, the error in the pixels where the measurement is possible becomes large (see Fig. \ref{fig_variance-plot}). The datapoints spread out over a wide range in the $N$(C$^{18}$O) vs. $A_\mathrm{V}$ correlation diagram (Fig. \ref{fig_correlations}d). The linear fit to the datapoints follows the Eq. \ref{eq_result-in-chaI} very closely.

\begin{figure*}
   \centering
    \includegraphics[width=1.25\columnwidth]{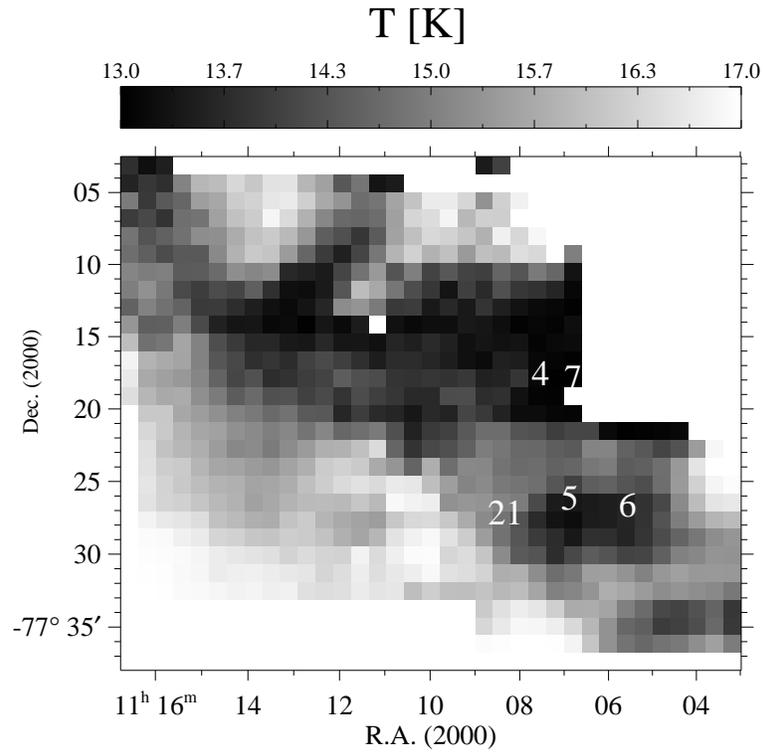}
      \caption{A tentative dust temperature map of Cha I derived from the ISO 100, 150 and 200 $\mu$m observations (Lehtinen et al. in prep.). The numbers refer to some of the C$^{18}$O clumps identified by H05. The temperature minimum coincides with the region of the clumps 4 and 7, revealing a prominent cold, heavily depleted core.
              }
         \label{fig_temperature-map}
   \end{figure*}

The westernmost C$^{18}$O maximum including clumps 4 and 7 shows a peculiar behaviour with close to constant $N$(C$^{18}$O) level over the range $A_\mathrm{V}\approx 5-14^m$. This results in a narrow, almost horizontal relation implying that the level of depletion in this clump is particularly high (Fig. \ref{fig_correlations}e). A few identified YSOs lie loosely scattered in this region, but generally it belongs to the low-activity part of the complex. The high level of depletion, however, is an implication of large density and low temperature. Fig. \ref{fig_temperature-map} showing a tentative dust temperature map derived from the ISO 100, 150 and 200 $\mu$m observations (Lehtinen et al. \cite{lehtinen05}, in preparation) indeed points out that this region represents the coldest region of the southern part of the cloud. We thus identify the region as a prominent, cold, heavily depleted dense core.

The southernmost of the large C$^{18}$O maxima, consisting of clumps 5 and 6 suffers from the same lack of stars due to high extinction as the central region. The maximum extinction derived is about $14^m$, but taking the region without any background stars into account, the maximum extinction is substantially higher making this region most likely the densest part of the southern side of the Cha I cloud. This clump also corresponds to the position of a very cold core detected by T\'oth et al. (\cite{toth00}) from the ISOPHOT Serendipity Survey data, having an ISO $I$(170)/$I$(100) color temperature $\lesssim$ 12.5 K. The Fig. \ref{fig_temperature-map} confirms that the clumps indeed coincide with the local temperature minimum.

For comparison, a pair of clumps with very different $N$(C$^{18}$O) vs. $A_\mathrm{V}$ relations are shown in Figs. \ref{fig_correlations}h-i. Spatially the clumps belong to the southern part of Cha I and are separated only about 15$'$. The former one has only very weak counterpart in C$^{18}$O emission, even though its extinction rises above $A_\mathrm{V}$=10$^m$. The latter one, located close to a group of identified YSOs has a much stronger C$^{18}$O emission compared to the extinction.

On the basis of the earlier discussion, it seems 
likely that the clumps where the $N({\rm
C^{18}O})$ vs. ${A_\mathrm{V}}$ proportionality is shallow (Figs. \ref{fig_correlations}e and h) are denser than those which roughly follow the average
relationship (e.g. Fig. \ref{fig_correlations}g). The latter can represent column
density maxima without well-defined centres of gravity. On the other hand, in the two clumps in the northern part of Cha I
(Fig. \ref{fig_correlations}c) the $N({\rm C^{18}O})/{A_\mathrm{V}}$ ratio is larger that the
average. This region has recently formed a dense cluster of stars, and is
associated with an extented bipolar outflow (Mattila et al. \cite{mattila89}).  It
seems therefore possible that the relatively large abundance of
gaseous C$^{18}$O in these clumps is a result of intensified desorption
related to heating by shocks or radiation from newly born stars.


\section{Conclusions}
\label{sec_conclusions}

We have derived the maps of visual extinction in the active stellar star forming cloud Chamaeleon I and in a dense quiescent filament of the Chamaeleon III cloud using the NICER color excess technique and 2MASS data. We have studied the $N$(C$^{18}$O) / A$_\mathrm{V}$ ratio and searched for differences between the two clouds, and between the regions and clumps in Chamaeleon I. For this study we have used the C$^{18}$O data from the SEST telescope published in Haikala et al. (\cite{haikala05}, Cha I) and Gahm et al. (\cite{gahm02}, Cha III), by smoothing them to the resolution of the extinction maps. We have also used 2MASS data to derive the reddening law in Chamaeleon I. The conclusions of our work are summarized in the following:

\begin{enumerate}

\item The derived visual extinction maps of Cha I and Cha III-B resemble well the complex column density structure seen in the C$^{18}$O maps. With the 2$'$ resolution the maximum extinction reaches to $A_\mathrm{V} \approx 19^m$ in Cha I and $A_\mathrm{V} \gtrsim 10^m$ in Cha III. Several opaque spots with $A_\mathrm{V} \gtrsim 10^m$ can be found on the Cha I map, and most of them can be attributed to the C$^{18}$O clumps identified by Haikala et al. (\cite{haikala05}).

\item The average relationship between $N$(C$^{18}$O) and $A_\mathrm{V}$ is similar in Cha I
and Cha III, i.e. $N$(C$^{18}$O) $\sim 2 \times 10^{14}$ cm$^{-2} (A_\mathrm{V}-2$). This is in
contrast with the previous result of Hayakawa et al. (\cite{hayakawa01}), who find
a steeper proportionality in Cha I using NANTEN observations. The
reason for this discrepancy is probably related to an inexplicable
difference between the intensity scales of the two C$^{18}$O data sets
towards Cha I.

\item The color-color diagram ($H-K$, $J-H$) of 2MASS stars in Chamaeleon I reveals four new YSO candidates with NIR excess emission. The new sources with excess emission are listed in Table \ref{tab_new-ysos}.

\item Regional variations of the $N$(C$^{18}$O) / A$_\mathrm{V}$ ratio are found within Chamaeleon I. The most obvious example is the northernmost active star formation region where the $N$(C$^{18}$O) level is enhanced by a factor of $1.3 \pm 0.2$ with respect to the average relation in the cloud, and by a factor of 1.5-3.5 with respect to the clumps with C$^{18}$O level lower than average. We suggest that this is caused by an increased amount of gas phase C$^{18}$O as a result of active star formation. On the other hand, regions with a markedly shallow  $N$(C$^{18}$O) vs. A$_\mathrm{V}$ proportionality can be identified. These are likely to represent cold, dense cores with a high degree of molecular depletion.  It looks thus conceivable to use the combinaton of 2MASS
data and large scale C$^{18}$O surveys for finding low-temperature
molecular cores. This suggestion is substantiated by the fact that the
most extensive of the cores standing out on the $N$(C$^{18}$O) vs. A$_\mathrm{V}$ plot coincides with a local dust temperature minimum discovered using ISO data.

\end{enumerate}


\begin{acknowledgements}

We thank the referee, E. Bergin, for useful comments and suggestions. This publication makes use of data products from the Two Micron All Sky Survey, which is a joint project of the University of Massachusetts and the Infrared Processing and Analysis Center/California Institute of Technology, funded by the National Aeronautics and Space Administration and the National Science Foundation. J.K. acknowledges the support of the Academy of Finland, grant Nr. 206049.

\end{acknowledgements}



\newpage

\newpage

\pagebreak

\end{document}